\documentclass[fleqn,usenatbib]{mnras}

\usepackage{newtxtext,newtxmath}

\usepackage[T1]{fontenc}

\DeclareRobustCommand{\VAN}[3]{#2}
\let\VANthebibliography\thebibliography
\def\thebibliography{\DeclareRobustCommand{\VAN}[3]{##3}\VANthebibliography}

\usepackage{ae,aecompl}
\usepackage{float}
\usepackage{soul}
\usepackage[usenames,dvipsnames]{xcolor}
\usepackage{subcaption}
\usepackage{tikz}
\usetikzlibrary{decorations.pathmorphing}
\usepackage{graphicx}	
\usepackage{amsmath}	

\definecolor{electricviolet}{rgb}{0.56, 0.0, 1.0}

\title[Galaxy and Halo sizes]{The relationship between galaxy and halo sizes in the Illustris and IllustrisTNG simulations}

\author[Karmakar et al.]{
Tathagata Karmakar${^{1,2,3}}$\thanks{tkarmaka@UR.Rochester.edu}
Shy Genel$^{4,5}$,
Rachel S. Somerville$^{4}$
\\
$^{1}$Department of Physics and Astronomy, University of Rochester, Rochester, NY 14627, USA.\\
$^{2}$Center for Coherence and Quantum Optics, University of Rochester, Rochester, NY 14627, USA.\\
$^{3}$Institute for Quantum Studies, Chapman University, Orange , CA 92866, USA.\\
$^{4}$Center for Computational Astrophysics, Flatiron Institute, 162 5th Ave, New York, NY 10010, USA\\
$^{5}$Columbia Astrophysics Laboratory, Columbia University, 550 West 120th Street, New York, NY 10027, USA\\}

\pubyear{0000}

\begin{document}
\label{firstpage}
\pagerange{000--000}
\maketitle

\begin{abstract}
Abundance matching studies have shown that the average relationship between galaxy radius and dark matter halo virial radius remains nearly constant over many orders of magnitude in halo mass, and over cosmic time since about $z=3$. In this work, we investigate the predicted relationship between galaxy radius $r_{e}$ and halo virial radius $R_{\rm h}$ in the numerical hydrodynamical simulations Illustris and IllustrisTNG from $z\sim 0$--3, and compare with the results from the abundance matching studies. We find that Illustris predicts much higher $r_e/R_{\rm h}$ values than the constraints obtained by abundance matching, at all redshifts, as well as a stronger dependence on halo mass. In contrast, IllustrisTNG shows very good agreement with the abundance matching constraints. In addition, at high redshift it predicts a strong dependence of $r_e/R_{\rm h}$ on halo mass on mass scales below those that are probed by existing observations. We present the predicted $r_e/R_{\rm h}$ relations from Illustris and IllustrisTNG for galaxies divided into star-forming and quiescent samples, and quantify the scatter in $r_e/R_{\rm h}$ for both simulations. Further, we investigate whether this scatter arises from the dispersion in halo spin parameter and find no significant correlation between $r_e/R_{\rm h}$ and halo spin. We investigate the paths in $r_e/R_{\rm h}$ traced by individual haloes over cosmic time, and find that most haloes oscillate around the median $r_e/R_{\rm h}$ relation over their formation history.
\end{abstract}




\section{Introduction}
\label{sec:intro}
In the modern paradigm for galaxy formation, galaxies form within dark matter haloes, overdense gravitationally bound dark-matter dominated structures within which gas accretes and cools \citep{SD15}.  The \emph{virial radius} of a dark matter halo is typically defined, somewhat arbitrarily, as the radius within which the mean overdensity with respect to the background (or critical) density of the Universe\footnote{Conventions regarding halo definitions vary; see Section~\ref{sec:sham} for details.} is some factor $\Delta_{\rm vir}$. As the gas cools down sufficiently and condenses in the inner parts of halos, stars form and galaxies grow. Although the formation of dark matter haloes is fairly well understood and quantified through numerical simulation, our understanding of the baryonic processes that shape galaxy properties, such as star formation, stellar feedback, mergers, black hole growth and Active Galactic Nucleus (AGN) feedback, remains very incomplete \citep{SD15,NO17}. While the physics of galaxy formation seems exceedingly complex, the Universe provides us with clues to its underlying simplicity in the form of \emph{scaling relations}. These are tight observed relationships between galaxy global properties, such as stellar mass and metallicity, or global and structural properties, such as stellar mass and galaxy radius. The latter is the underlying focus of this study.

\emph{Abundance matching} is a well established empirical technique for constraining the relationship between galaxy and dark matter halo properties, within the paradigm in which galaxies are always hosted by DM haloes or sub-haloes \citep[for a recent review see][]{Wechler_tinker:2018}. Sub-haloes are haloes that have merged into and are orbiting within a larger virialized halo. In traditional Sub-halo Abundance Matching (SHAM), one derives a mapping between galaxy stellar mass (or luminosity) and dark matter halo mass ($m_*/M_{\rm h}$), using the (sub)halo mass function predicted by N-body simulations and an observed stellar mass (luminosity) function \citep[e.g.][]{moster:2010,behroozi:2010,rodriguez-puebla_galaxy-halo_2017}. In \emph{structural} SHAM, one takes this a step further by deriving a mapping between DM halo radius and galaxy radius, by adopting a SMHM relation (with some dispersion) and requiring that the model matches the observed size-mass relation. In this way, \citet{kravtsov:2013} showed that the relationship between galaxy size and halo virial radius ($r_e/R_h$) is consistent with a nearly constant average value across many orders of magnitude in mass and for galaxies of diverse morphology, from giant ellipticals to low mass dwarfs. Several later studies expanded this approach to high redshift \citep{shibuya:2015,Huang:2017,somerville_relationship_2018}. In particular, \citet[][hereafter S18]{somerville_relationship_2018} carefully combined observations from $z\sim 0$ with high redshift observations in a consistent manner, and also quantified the conditional size \emph{distributions}, as well as the mean $r_e/R_h$ relation. S18 showed that the nearly linear relation between $r_e$ and $R_h$ holds up to $z\sim 3$, with only weak evolution in the value of the ratio $r_e/R_h$. We focus on the S18 abundance matching results in this work, but they are broadly consistent with those of other studies in the literature. 

The results of the structural abundance matching studies mentioned above were unexpected, and prompted many open questions, including the following. 1) Why should the ratio between galaxy size and halo size be close to constant over many orders of magnitude in halo mass and many billions of years in cosmic time? This is especially mysterious given that the halo virial radius is in many ways an arbitrary quantity. 2) What can we learn from the empirical constraints on $r_e/R_h$ about the physical processes that shape the structural properties of galaxies? 3) What is the origin of the \emph{dispersion} in $r_e/R_h$ at fixed halo mass\footnote{Recall that for any given overdensity-based definition of the virial radius, virial mass and radius are perfectly interchangeable as there is zero scatter in the relationship between them.}? Can this be connected to a second or higher order halo parameter beyond mass? 4) What kind of paths in $r_e/R_h$ space do galaxies trace over cosmic time? Does a given object tend to always be high or low relative to the average $r_e/R_h$ over its whole history, or do galaxies oscillate around the mean $r_e/R_h$ relation?

In the classical picture of disc formation, dark matter haloes acquire angular momentum via quadrupole interactions with neighboring proto-galaxies \citep{peebles_origin_1969}.  One can define a dimensionless spin parameter for a halo with angular momentum J, energy E and mass M \citep[see][]{peebles_origin_1969}: \begin{equation}
    \begin{split}
        \lambda= 
        \frac{J |E|^{1/2}}{G M^{5/2}}
    \end{split}
    \end{equation} 
     which describes how rotationally supported the halo is. 
Analytic models employing the angular momentum partition ansatz assume that hot gas is ``spun up'' during the halo formation, and therefore has the same specific angular momentum as the DM halo, and that the specific angular momentum of the gas is conserved as it collapses to form a disc. Under the simplest set of assumptions, one expects the size of a disc forming under these conditions to obey a proportionality with the product of the halo spin and the halo radius: $r_e \propto \lambda R_h$ \citep[][]{mo_formation_1998,dalcanton:1997,somerville_explanation_2008}. 
Numerous refinements to the simplest models have been developed in the literature, accounting for complications such as deviation of the halo gas profile from an isothermal sphere, modification of the inner halo profiles by gravity or energy input from stellar feedback, and transfer of angular momentum during the disc formation process \citep{blumenthal:1986,flores:1993,dutton:2007}. For a more detailed summary of these models, see S18 Section 5.4; however, we note that all of these models predict a fairly strong linear relationship between $r_e$ and $\lambda R_h$, modulated by other parameters such as the halo concentration and the disc baryon fraction. Moreover, S18 showed that the conditional size relation (the distribution of galaxy sizes in a given stellar mass bin) is remarkably similar to the distribution of $\lambda R_h$ in their SHAM, providing an indirect suggestion that the dispersion in galaxy size at fixed stellar mass \emph{is consistent with} being due to a proportionality between $r_e$ and $\lambda R_h$.

Modern numerical hydrodynamic simulations set within a cosmological context are a powerful tool to gain more insights into the physical processes behind the observed scaling relations. These simulations have demonstrated that galaxy internal structures, including size, are very sensitive to the details of the baryonic processes implemented in the simulations, in particular stellar and black hole feedback. Reproducing the observed size-mass relation for galaxies was a challenge for the earliest generations of numerical hydrodynamic simulations, which tended to produce galaxies that were too compact and bulge dominated \citep{sommer_larson:1999,navarro_steinmetz:2000,steinmetz_navarro:2002}. The more recent generation of cosmological hydrodynamic simulations have demonstrated quite good success at reproducing the observed size-mass relation and its evolution over cosmic time \citep{brooks_interpreting_2011,furlong_size_2017,genel_size_2018}. 

All current cosmological hydrodynamic simulations contain phenomenological ``sub-grid'' recipes to model processes that cannot be simulated explicitly, such as the formation of stars out of dense gas, the driving of galaxy-scale winds by massive stars and supernovae explosions, and accretion onto supermassive black holes \citep{SD15}. These sub-grid recipes contain adjustable parameters that are tuned to match a set of observations. Global properties, such as the galaxy stellar mass function, are often used for this purpose. It is particularly intriguing that a set of parameters that reproduces such global properties may not reproduce structural properties: for example, in the original Illustris simulations \citep{vogelsberger_model_2013}, which were tuned to match a set of global properties, the predicted galaxy sizes were about a factor of two larger than observations indicate \citep{Genel:2014}. In the updated IllustrisTNG simulations, the relationship between stellar mass and halo mass remains very similar to that in the original Illustris simulations, but the predicted sizes are significantly smaller \citep{Pillepich:2018,genel_size_2018}. Although it is now well established that galaxy sizes are sensitive to both stellar and AGN feedback, many of the details of how physical processes like feedback shape galaxy sizes remain unclear. \par
 
 
 In this paper, we use the Illustris and IllustrisTNG simulations as laboratories to study the relationship between galaxies and their host dark matter haloes. Although many works have studied the predictions of numerical simulations for the observed galaxy size-mass relation, as noted above, there has been little work quantifying the relationships between DM halo properties and galaxy properties, such as the stellar mass vs. halo mass relation and galaxy size vs. halo size  relation, in physics-based simulations. We compare the stellar mass vs. halo mass and galaxy size vs. halo size relations extracted from Illustris and IllustrisTNG with the abundance matching results of S18 over a range of cosmic time ($z\sim 3$--0). We quantify the galaxy-size vs. halo size relation for star-forming and quiescent galaxies in the simulations. Abundance matching techniques have a limited ability to constrain the \emph{dispersion} in these relationships, which can be measured directly in the simulations. This result provides an important input for empirical models. In addition, we attempt to understand the \emph{physical origin} of the dispersion in these relations, by examining whether the dispersion in size at fixed mass is correlated with halo spin in the simulations. We further explore the origin of the dispersion by tracing the growth of individual galaxies over cosmic time. \par
 
In section~\ref{sec:sham}, we present a brief summary of the abundance matching method and results of S18. In section~\ref{sec:sims} we provide a brief background on the Illustris and IllustrisTNG simulations. We present our main results in section~\ref{sec:results}. We discuss the interpretation of our results, and their relationship with previous results in the literature in section~\ref{sec:discussion}, and summarize our key findings and conclude in section~\ref{sec:conclusions}.

\section{Sub-halo abundance matching method and results}
\label{sec:sham}
    
In the simplest version of (sub)-halo abundance matching, one takes a population of dark matter haloes from a dissipationless N-body simulation and orders them by mass. One then takes a population of galaxies from an observational sample, orders them by the global property (such as stellar mass or luminosity), and assumes that the most massive (luminous) galaxy occupies the most massive halo, the second most massive galaxy occupies the second most massive halo, etc. This procedure, however, does not account for possible dispersion in the stellar mass vs. halo mass (SMHM) relation. More recent works instead fit for the parameters in a function describing the mean SMHM relation ($\langle m_*(M_h) \rangle$) in order to minimize the deviation between the model stellar mass function and an observed stellar mass function \citep{rodriguez-puebla_galaxy-halo_2017,moster:2018,behroozi:2019}. The dispersion in $m_*$ as a function of $M_h$ and redshift $z$ is also parameterized, and these parameters are constrained in the modeling procedure. Most SHAM studies also fit for galaxy clustering properties, which provide some constraints on the dispersion in $m_*(M_h)$. S18 adopt the SMHM relation and stellar mass dispersion relations from \citet{rodriguez-puebla_galaxy-halo_2017}.  Note that both S18 and \citet{rodriguez-puebla_galaxy-halo_2017} adopt the definition of halo mass and virial radius $M_{\rm vir} = \frac{4}{3} \Delta_{\rm vir} \rho_{\rm crit} R^3_{\rm vir}$, where $\Delta_{\rm vir}$ is given by Eqn.~6 of \citet{bryan:1998} and $\rho_{\rm crit}$ is the critical density of the Universe. 

S18 then used a ``forward modeling'' approach to constrain the median relationship between galaxy size and halo size as follows. They created a mock catalog based on the Bolshoi-Planck dissipationless N-body simulation \citep{rodriguez-puebla:2016} by assigning a stellar mass to each halo and sub-halo, including scatter as described above. They then binned their sample in stellar mass, and found the median halo radius in each bin. They then computed the median observed galaxy size in the same stellar mass bin. In this way, they obtained the median relation $r_e/R_h$. Other studies, such as \citet{kravtsov:2013} and \citet{Huang:2017}, have carried out a similar analysis using somewhat different techniques (see S18 for an extensive discussion of the differences), and obtained qualitatively similar results for $r_e/R_h$.  In ``backwards modeling'',  halo properties corresponding to an observational sample of galaxies are obtained by inverting a SHMR relation. As the SHMR becomes shallower at high halo masses, a positive deviation in stellar mass causes a larger deviation in halo mass than a comparable negative deviation does. Thus, in the presence of  scatter in the SHMR, ``backwards modeling'' leads to an overestimate in halo mass and sizes, making ``forward modeling'' a more reliable strategy (see S18 for a detailed discussion). 

The observational samples used to obtain the size-mass relations adopted in S18 come from DR2 of the Galaxy And Mass Assembly Survey (GAMA; \citealp{liske:2015}), which is an optically selected survey of nearby galaxies ($0.01 < z < 0.12$), and the CANDELS survey \citep{grogin:2011,koekemoer:2011}, which is based on observations with the WFC3 camera on the Hubble Space Telescope and probes galaxies out to high redshift ($z\sim 3$ in the study of S18). In both cases, stellar masses are estimated from the photometry using standard techniques. Also in both cases, the observed projected semi-major axis half-light radii in a fixed rest-frame band ($r_{\rm e,obs}$) from the studies of \citet{lange:2015} and \citet{vanderwel:2014} are converted to 3D half stellar mass radii ($r_{\star,\mathrm{3D}}$) using  the formula $r_{\rm e,obs}=f_pf_kr_{\star,\mathrm{3D}}$. Here $f_p$ accounts for the projection from 3D to 2D, and $f_k$ accounts for between conversion  half-light radii and half stellar mass radii. In S18, a value of $(f_pf_k)_\mathrm{disk} = 1.2$ has been adopted for disk galaxies (i.e.~galaxies with S\'ersic index $n_s<2.5$), while a value of $(f_pf_k)_\mathrm{spheroid} = 0.78$ has been adopted for spheroids ($n_s>2.5$). Since $f_p$ depends on galaxy shapes, it has the potential to introduce additional stellar mass and redshift dependence in $r_{\rm e,obs}/r_{\star,\mathrm{3D}}$. The values  of $f_k$  are found to lie between $1.12-1.5$ and do not show any significant dependence on galaxy mass or redshift (see \cite{lange:2015} and \cite{Wuyts_2012}). See S18 for a more detailed discussion of the projected light to 3D stellar mass size conversion procedure and references for the adopted parameter values.

S18 and \citet{kravtsov:2013} found the striking result that the average relationship between galaxy size and halo virial radius $r_e/R_h$ at $z=0$ \emph{remains nearly constant over several orders of magnitude in halo mass}. This result is particularly surprising given that the galaxy population is dominated by disc-like galaxies at lower stellar (halo) masses (see Figure 4 in S18), but dominated by elliptical type galaxies at higher masses, and these objects are thought to form in very different ways. S18 and \citet{Huang:2017} further showed that this average ratio $r_e/R_h$ remains nearly independent of halo mass out to high redshift $z\sim 3$, and its value evolves only mildly with redshift over this period of more than 10 billion years in cosmic time. 

These studies have left some open questions. First, some studies (\citet{kravtsov:2013} and \citet{Huang:2017}) present results for $r_e/R_h$ for galaxies dis-aggregated into disc-dominated and spheroid dominated types. However, this relies on the assumption that the SMHM relation for these different types of galaxies is the same, which may not be the case \citep{Cui:2021}. Second, none of these studies are able to constrain the \emph{dispersion} in $r_e/R_h$. Third, the physical reason that all types of galaxies over such a large range in halo mass and redshift should obey this nearly universal relationship between the size of their stellar body and the virial radius of their dark matter halo remains unclear.

\section{The simulated galaxy populations}
\label{sec:sims}

Illustris \citep{Genel:2014,vogelsberger_introducing_2014,vogelsberger_properties_2014} and IllustrisTNG \citep{marinacci_first_2018,naiman_first_2018,nelson_first_2018a,pillepich_first_2018,springel_first_2018} are suites of hydrodynamical simulations of galaxy formation and evolution in large cosmological volumes. Here we focus on the highest-resolution levels of the $(\sim100~\text{Mpc})^3$ volume from each of these suites, which we hereafter refer to simply as Illustris and IllustrisTNG. Both employ the quasi-Lagrangian code AREPO to evolve up to a total of $\sim 2\times1820^3$ resolution elements representing both dark matter and baryons in a $(75h^{-1}~\text{Mpc})^3$ volume. Dark matter is assumed to be collisionless and the N-body problem is solved using a TreePM method. For the hydrodynamical evolution, the volume is discretized using a Voronoi tessellation whose mesh-generating points move and (de)refine in a quasi-Lagrangian manner. Further, the simulation takes into account various physical processes that are believed to be critical for galaxy formation. These processes include radiative cooling, star formation after gas crosses a critical number density threshold of $0.13\text{cm}^{-3}$, metal enrichment through stellar evolution, feedback of gas into the circumgalactic medium via galactic winds, and the feedback associated with supermassive black holes (SMBH) at the centers of galaxies (see \citealp{vogelsberger_model_2013,weinberger:2017,Pillepich:2018}).\par

Despite showing good qualitative agreement with many observational results, some predictions from Illustris are in tension with observations, such as larger galaxy sizes compared to the observations, and overprediction of the number density of both high and low mass galaxies at $z\sim0$. IllustrisTNG attempts to address some of these issues by adopting a few key differences in some of the physical processes. These include modeling of magnetic fields (via the ideal MHD approximation), adding more flexibility to the galactic wind model, replacing the thermal bubble model for AGN feedback at low accretion rates with a kinetic AGN feedback model, increasing the black hole seed mass and removing the black hole accretion artificial boost factor. The cosmological parameters have also been slightly updated in IllustrisTNG ($\Omega_{m}$ = 0.3089, $\Omega_{\Lambda}$ = 0.6911, $\Omega_{b}$ = 0.0486, $\sigma_{8}$ = 0.8159 and $h$ = 0.6774) compared to Illustris ($\Omega_{m}$ = 0.2726, $\Omega_{\Lambda}$ = 0.7274, $\Omega_{b}$ = 0.0456, $\sigma_{8}$ = 0.809 and $h$ = 0.704).

The objects we refer to as galaxies in this work are identified using the SUBFIND algorithm \citep{springel:2001}, which finds self-bound objects around peaks in the density field. When referring to virial properties of the host haloes, we employ a definition corresponding to a spherical region around the most bound particle in the halo that contains a total matter overdensity of $\Delta_{\rm vir}$ (as given by Eqn.~6 of \citet{bryan:1998}) with respect to the critical density of the universe. Galaxy 3D sizes $r_e$ are defined as the radius of a sphere that contains half of the total stellar mass, while the standard stellar mass value we use for each galaxy is the mass of stars within twice its $r_e$. Since we focus on relations between galaxies and their host haloes, we only study central galaxies, and we focus on those with stellar masses larger than $10^8\,M_\odot$, which are well-resolved in the $(\sim100~\text{Mpc})^3$ volumes we employ. 

\begin{figure} 
\includegraphics[width=\columnwidth]{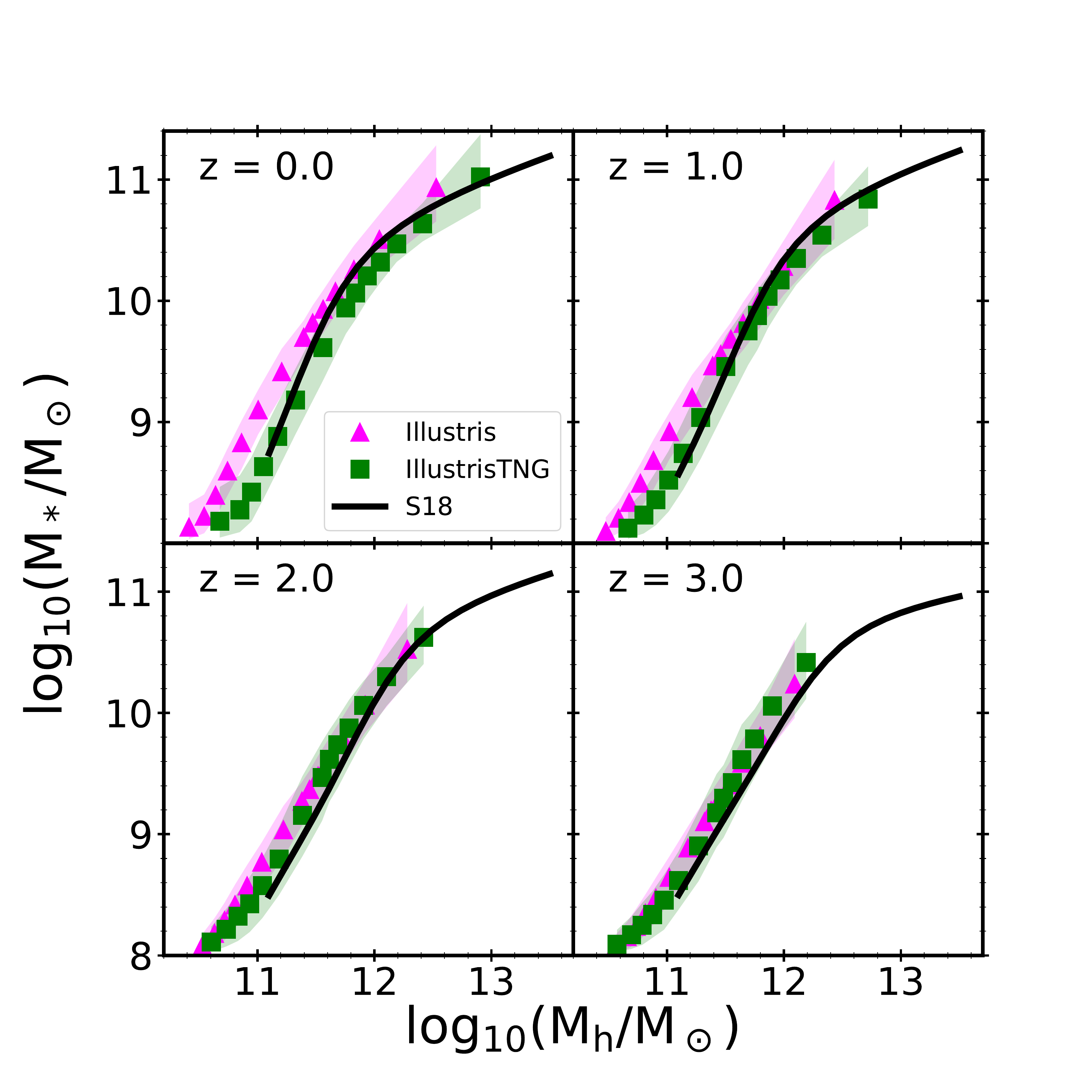}
\caption{Comparison of stellar-to-halo mass relations predicted by the Illustris and IllustrisTNG simulations (pink triangles and green squares, respectively) and that adopted in the SHAM of S18 (black lines), in several redshift bins from $z=0$ to 3 as indicated on the panels. The symbols show the medians from the simulations, and shaded areas show the 16th and 84th percentiles. The stellar to halo mass relations are very similar in the two simulations, and agree well with the SHAM results. }
\label{fig:Ms_Mh}
\end{figure}

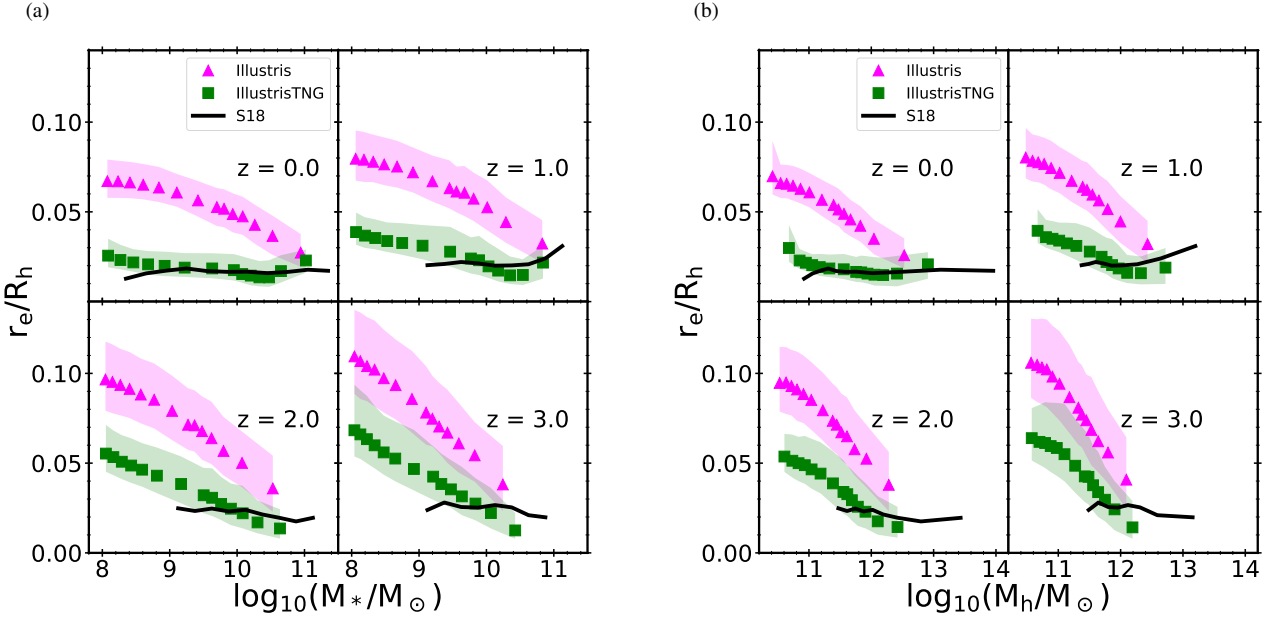
\begin{figure*}
    \includegraphics[width=0.49\textwidth]{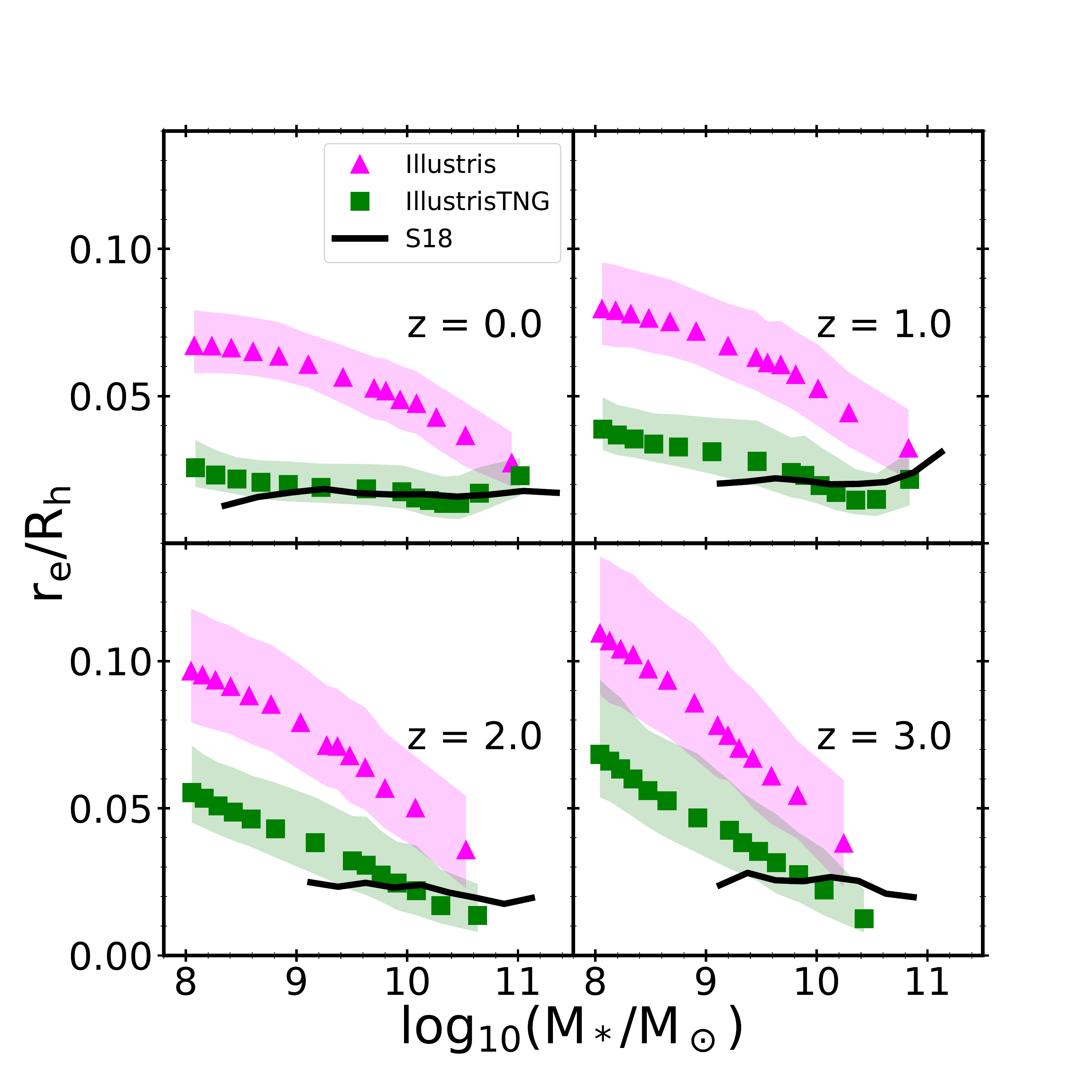}
    \includegraphics[width=0.49\textwidth]{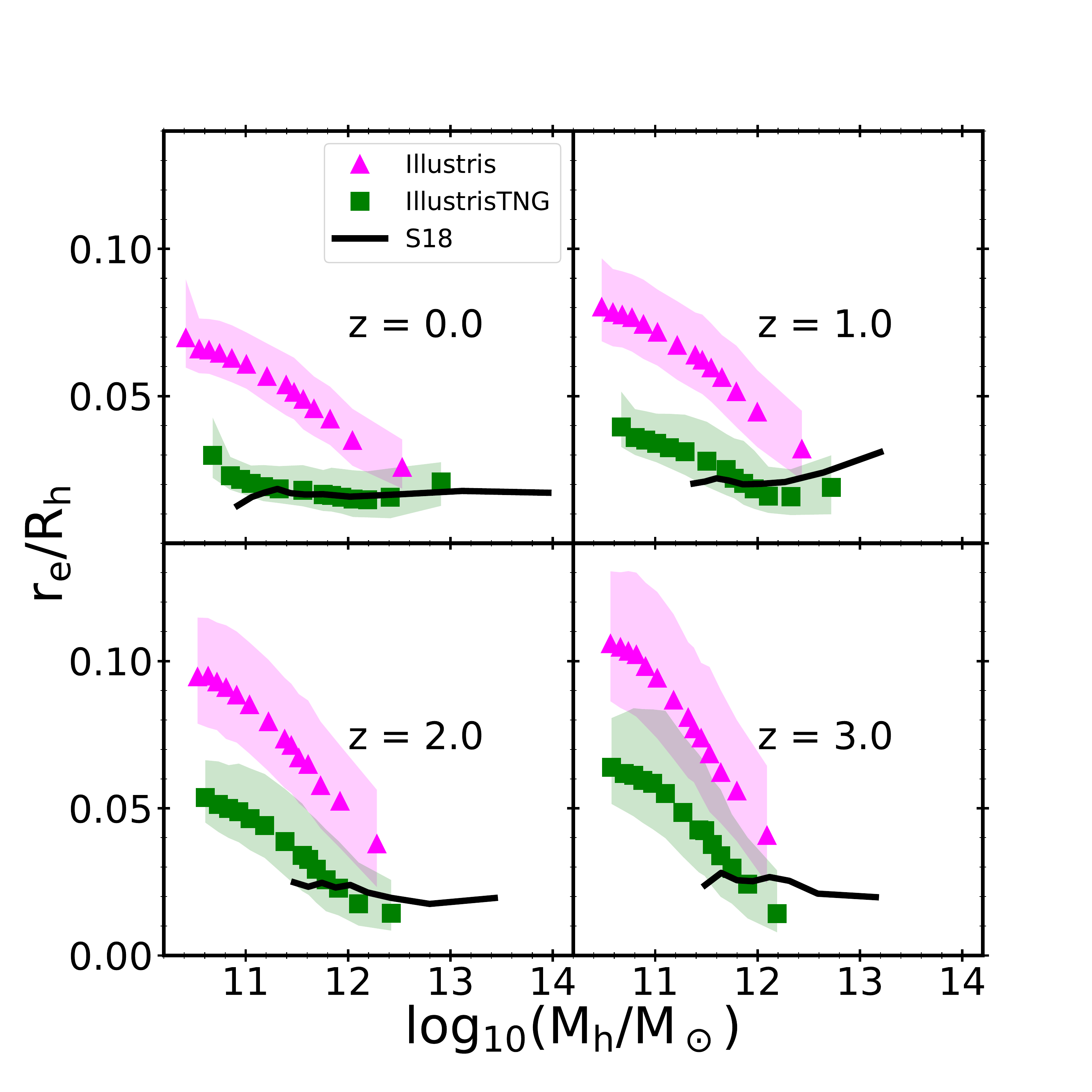}
    \begin{tikzpicture}[overlay]
	\node[] at (-17,8.2) {(a)};
	\node[] at (-8.2,8.2) {(b)};
	\end{tikzpicture}
    \caption{Galaxy radius to halo radius ratio $r_e/R_h$ of central galaxies versus stellar (left) and halo  mass (right). Magenta (green) symbols represent the medians from the Illustris (IllustrisTNG) simulation and corresponding shaded regions represent 16th and 84th percentiles. Black curves show the SHAM results from S18. At low redshifts, the IllustrisTNG $r_e/R_h$ relation is rather flat, and agrees well with the SHAM results. However, at high redshifts both simulations show a stronger dependence of $r_e/R_h$ on mass, in mild tension with the flat SHAM results.}
    \label{fig:rat_Mh}
\end{figure*}

\section{Results}
\label{sec:results}
\subsection{Comparison with Abundance Matching Results}
We start by comparing the median relationships between stellar mass and halo mass with the abundance matching results of S18 for both the Illustris and IllustrisTNG simulations. Figure \ref{fig:Ms_Mh} shows the stellar-halo mass relation (SHMR) of the two simulations compared with that adopted in S18. We note the very good agreement between, in particular, the IllustrisTNG predictions and the SHAM results at all redshifts. Although IllustrisTNG was tuned, in part, to match the stellar mass function at $z=0$, it was not explicitly tuned to match stellar mass functions at high redshift. At low redshift and low halo masses, the galaxy masses in Illustris are slightly higher at fixed halo mass than those in IllustrisTNG. This reflects the higher abundances of low mass galaxies predicted by the original Illustris simulation relative to observational constraints, which the revised physics in TNG was designed to mitigate. At $z>1$, the SMHM relations predicted by the two simulations are nearly indistinguishable. \par

Next we investigate the relationship between $r_e/R_h$ and stellar mass or halo mass. Figure \ref{fig:rat_Mh} shows $r_e/R_h$ vs. stellar mass on the left panel and $r_e/R_h$ vs. halo mass on the right panel. The IllustrisTNG simulation is consistent with the relatively constant behavior of $r_e/R_h$ with stellar and halo mass given by SHAM over the mass range where there are observational constraints. Interestingly, however, at lower masses and especially at high redshift, the simulations predict a significant increase in $r_e/R_h$ with decreasing mass. The original Illustris simulation not only predicts a higher normalization for $r_e/R_h$ than IllustrisTNG --- consistent with the well-known tendency of original Illustris to produce galaxies with sizes that are too large compared with observations --- but it also predicts a stronger dependence of $r_e/R_h$ on halo mass than IllustrisTNG. This is very interesting, as it suggests that the very flat behavior of $r_e/R_h$ with halo mass implied by the SHAM (and perhaps pertaining in the real Universe) may not be very generic. 

\begin{figure*}
\includegraphics[width=\linewidth]{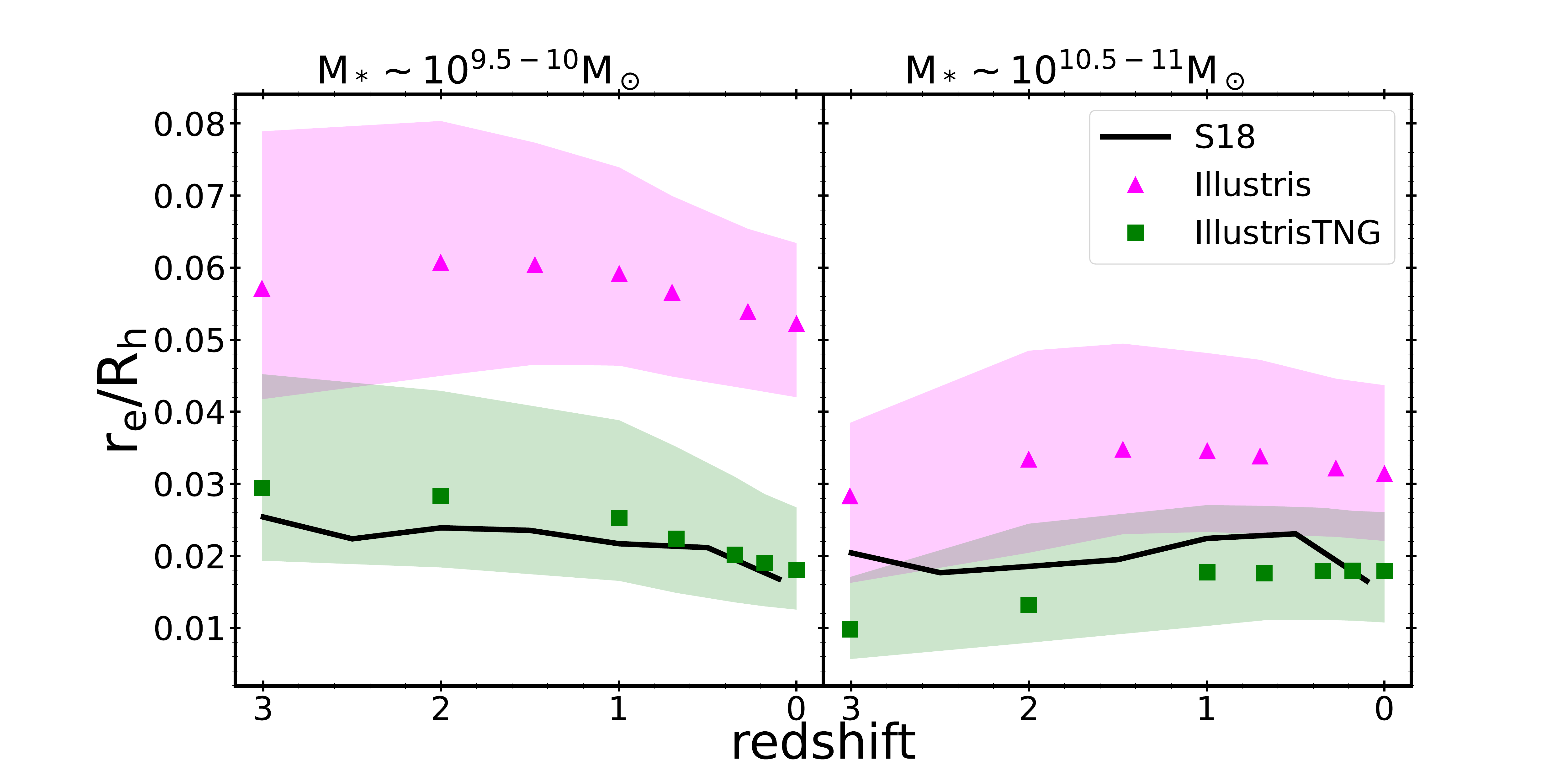}
\caption{Galaxy radius to halo radius ratio as a function of redshift for Illustris, IllustrisTNG and the S18 SHAM for stellar mass bins $10^{9.5-10}M_\odot$ (left) and $10^{10.5-11}M_\odot$ (right). All three models show weak, qualitatively similar redshift evolution.}
\label{fig:rat_redshift}
\end{figure*}

Figure~\ref{fig:rat_redshift} shows $r_e/R_h$ as a function of redshift for two different stellar mass bins ($10^{9.5}-10^{10}M_\odot$ on the left, $10^{10.5}-10^{11}M_\odot$ on the right) for both the Illustris and IllustrisTNG simulations, and the SHAM. The IllustrisTNG simulation shows qualitatively very similar behavior to the SHAM results, indicating a mild increase in $r_e/R_h$ with cosmic time for the lower mass bin, and a mild decrease for the higher mass bin. Once again, the normalization of $r_e/R_h$ is much higher at all redshifts for Illustris than for IllustrisTNG or the SHAM (in particular at lower masses), but the redshift dependence is similar. The dispersion in $r_e/R_h$ is also somewhat higher in the original Illustris simulation.

\begin{figure}
\begin{subfigure}{\linewidth}
\centering
\includegraphics[width=\linewidth]{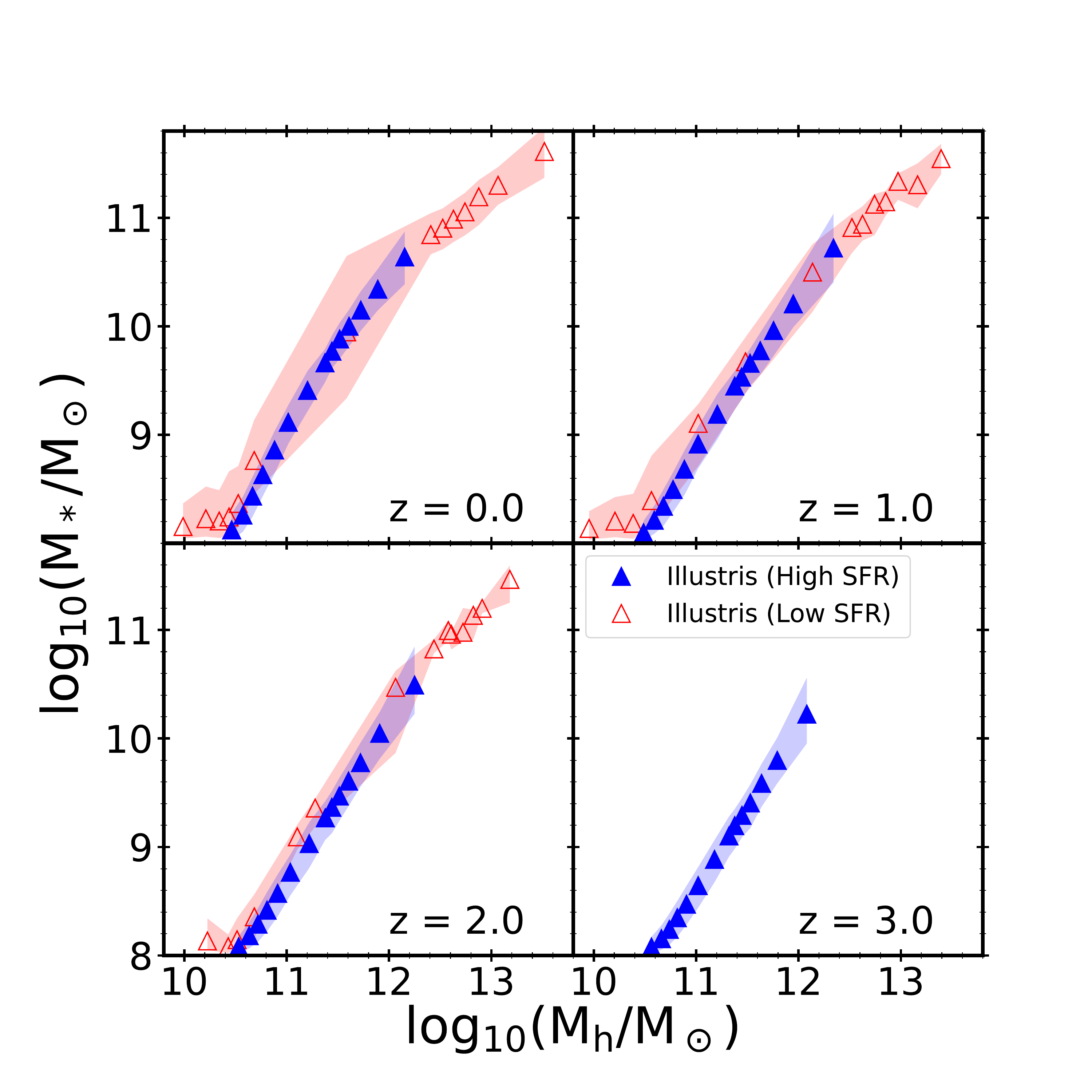}
\label{Ms_Mh_ill}
\end{subfigure}
\begin{subfigure}{\linewidth}
\centering
\vspace{-1cm}
\includegraphics[width=\linewidth]{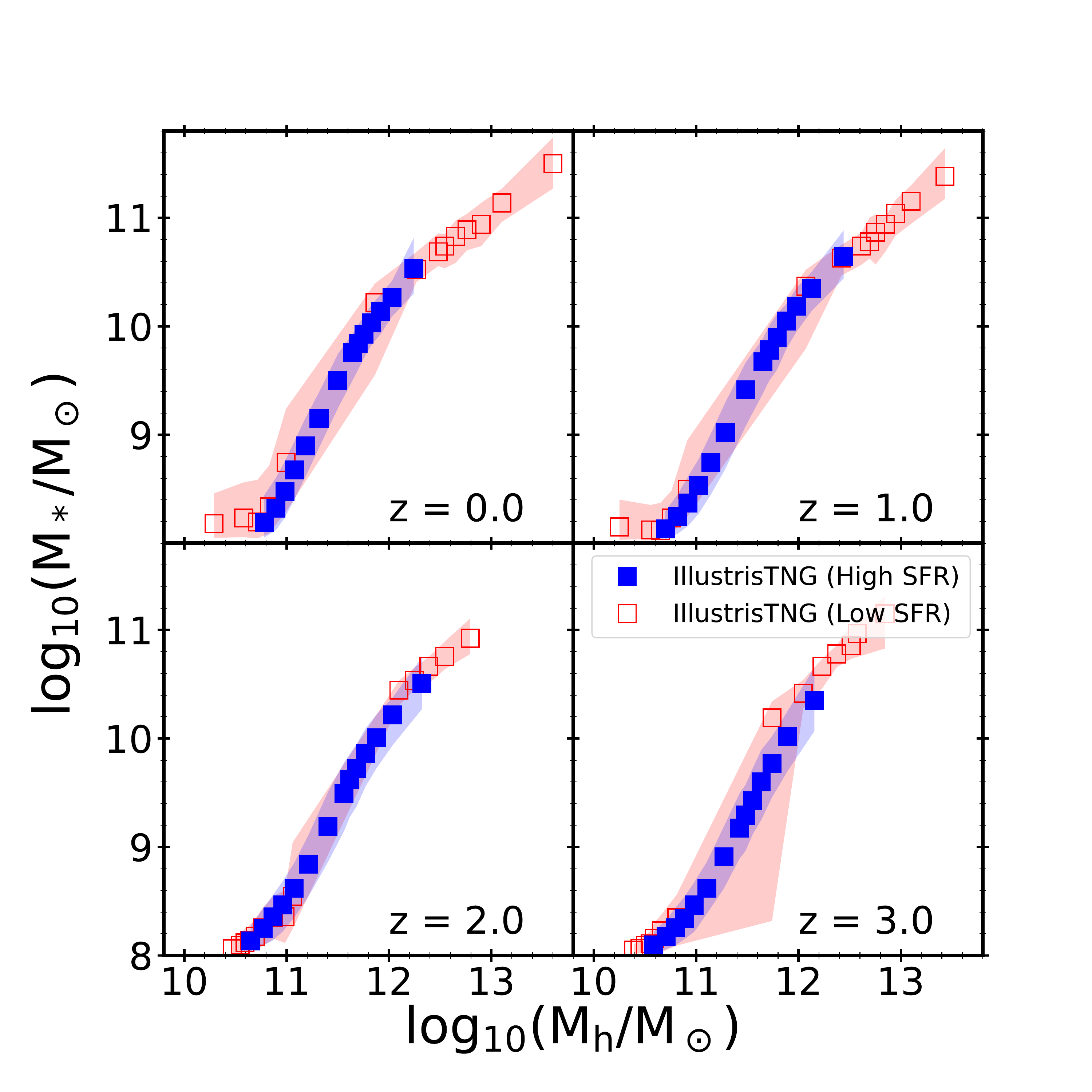}
\label{Ms_Mh_tng}
\end{subfigure}
\caption{Stellar-to-halo mass relation for Illustris (top) and IllustrisTNG (bottom). The red symbols represent low-SFR galaxies and the blue represent high-SFR ones. Corresponding shaded regions represent 16th to 84th percentiles. At $z=3$, for Illustris, low-SFR galaxies are not shown due to their extreme scarcity. In both simulations, the SMHM relation is very similar for high- and low-SFR galaxies. }
\label{fig:Ms_Mh_SFR}
\end{figure}

\begin{figure*}
\centering
\hspace{-0.2cm}
\begin{subfigure}{0.49\linewidth}
\includegraphics[width=\linewidth]{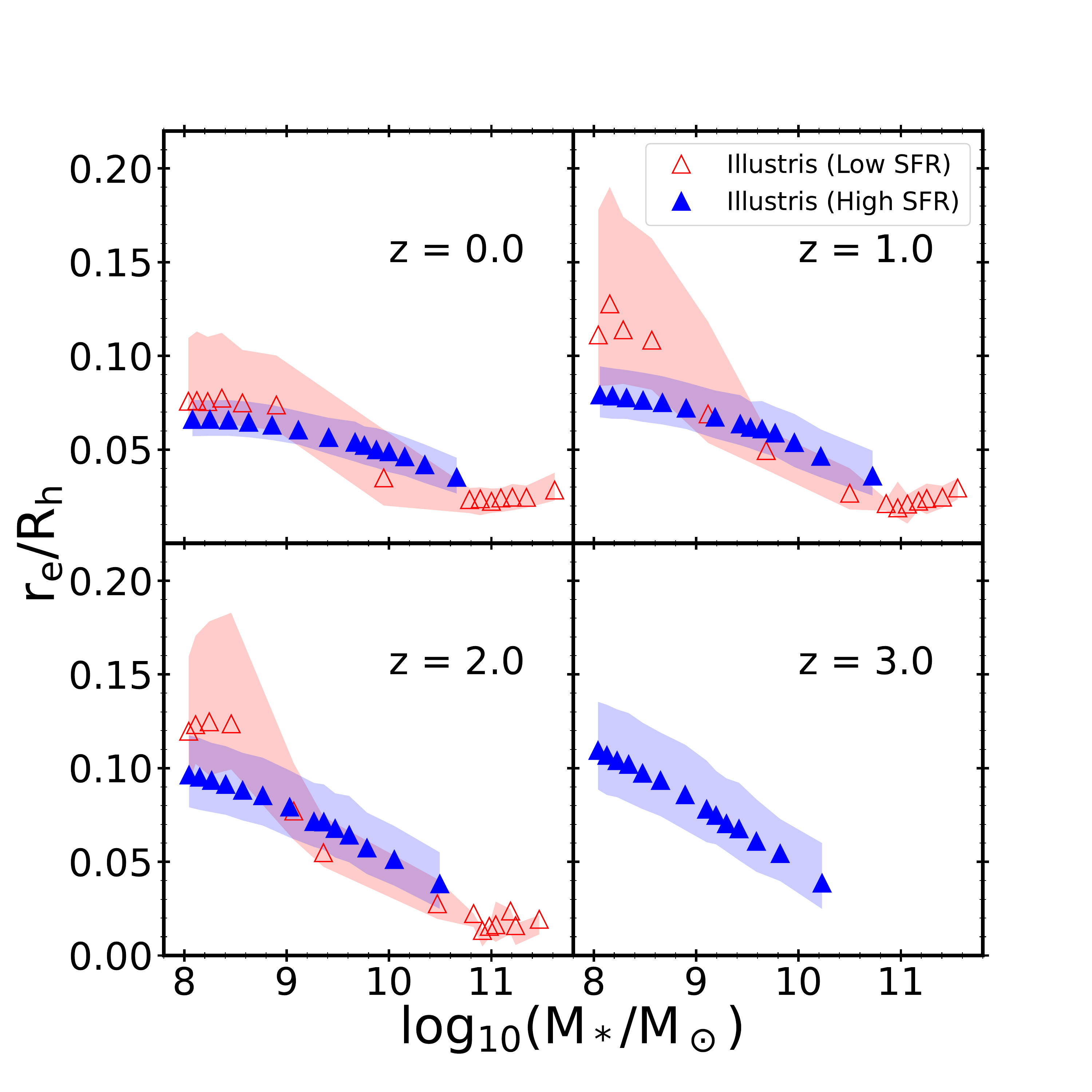}
\label{rat_s_ms_ill}
\end{subfigure}
\hspace{-0.8cm}
\begin{subfigure}{0.49\linewidth}
\includegraphics[width=\linewidth]{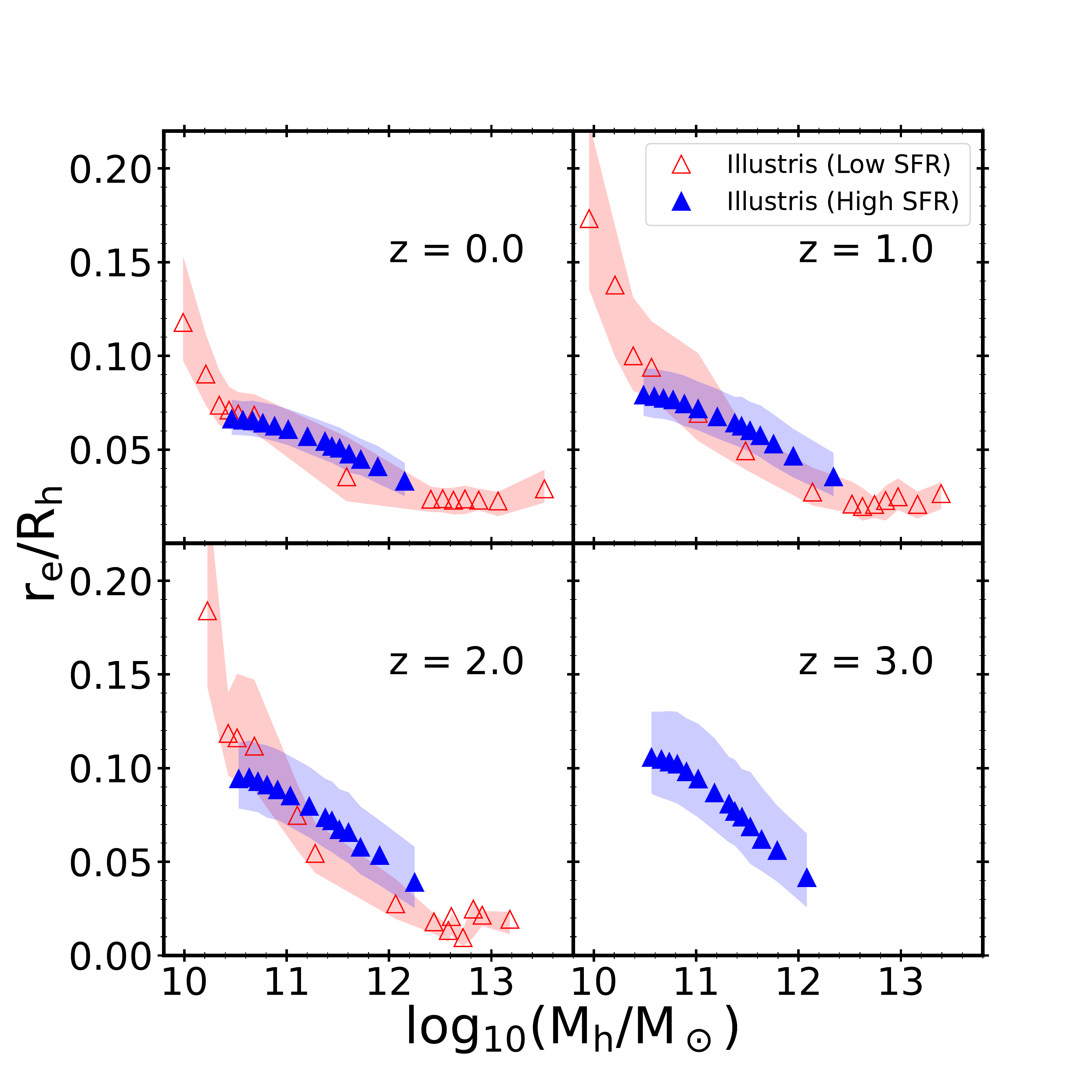}
\label{rat_s_mh_ill}
\end{subfigure}
\vspace*{-.1cm}
\hspace{0.1cm}
\begin{subfigure}{0.49\linewidth}
\vspace*{-.8cm}
\includegraphics[width=\linewidth]{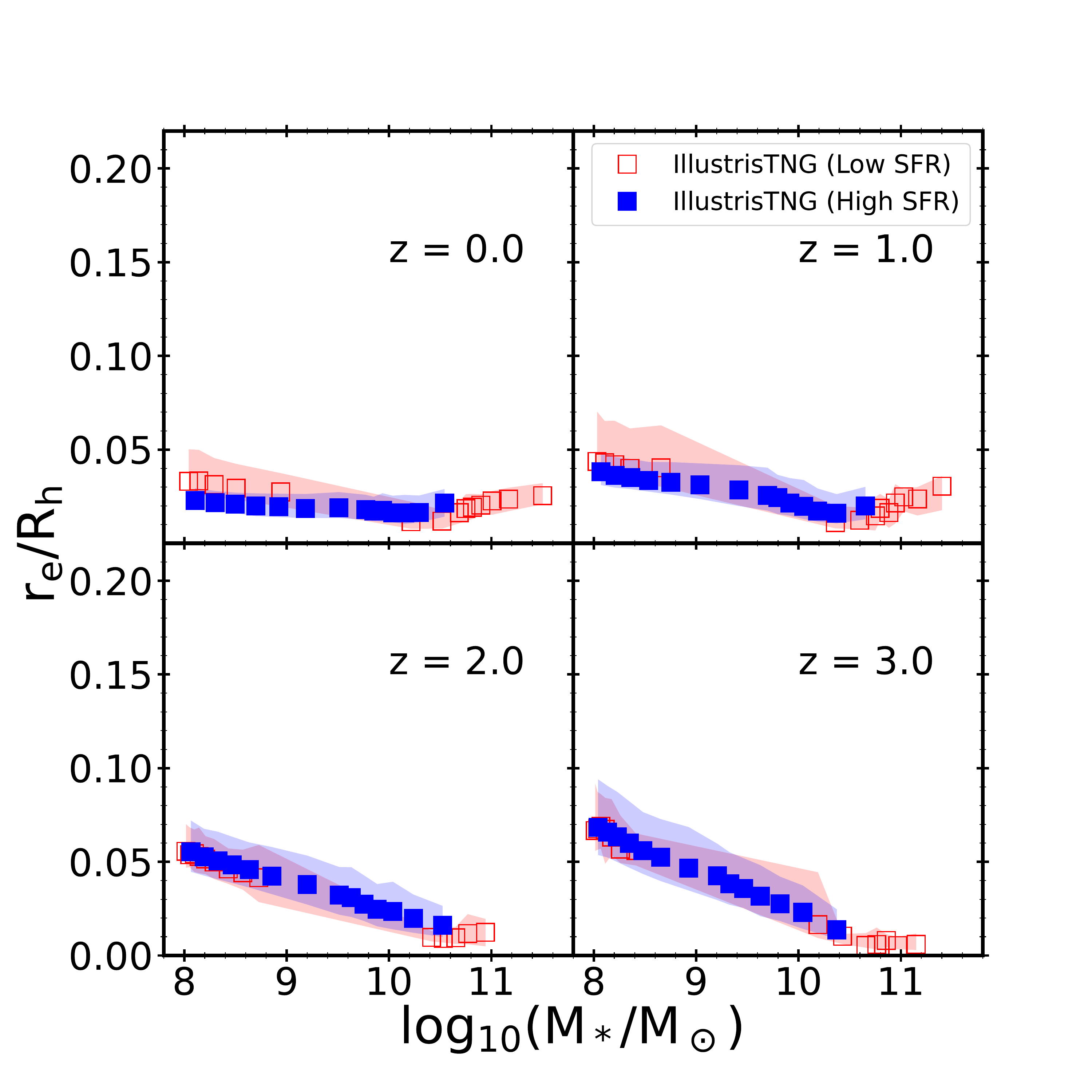}
\label{rat_s_ms_tng}
\end{subfigure}
\hspace{-0.8cm}
\begin{subfigure}{0.49\linewidth}
\vspace*{-.8cm}
\includegraphics[width=\linewidth]{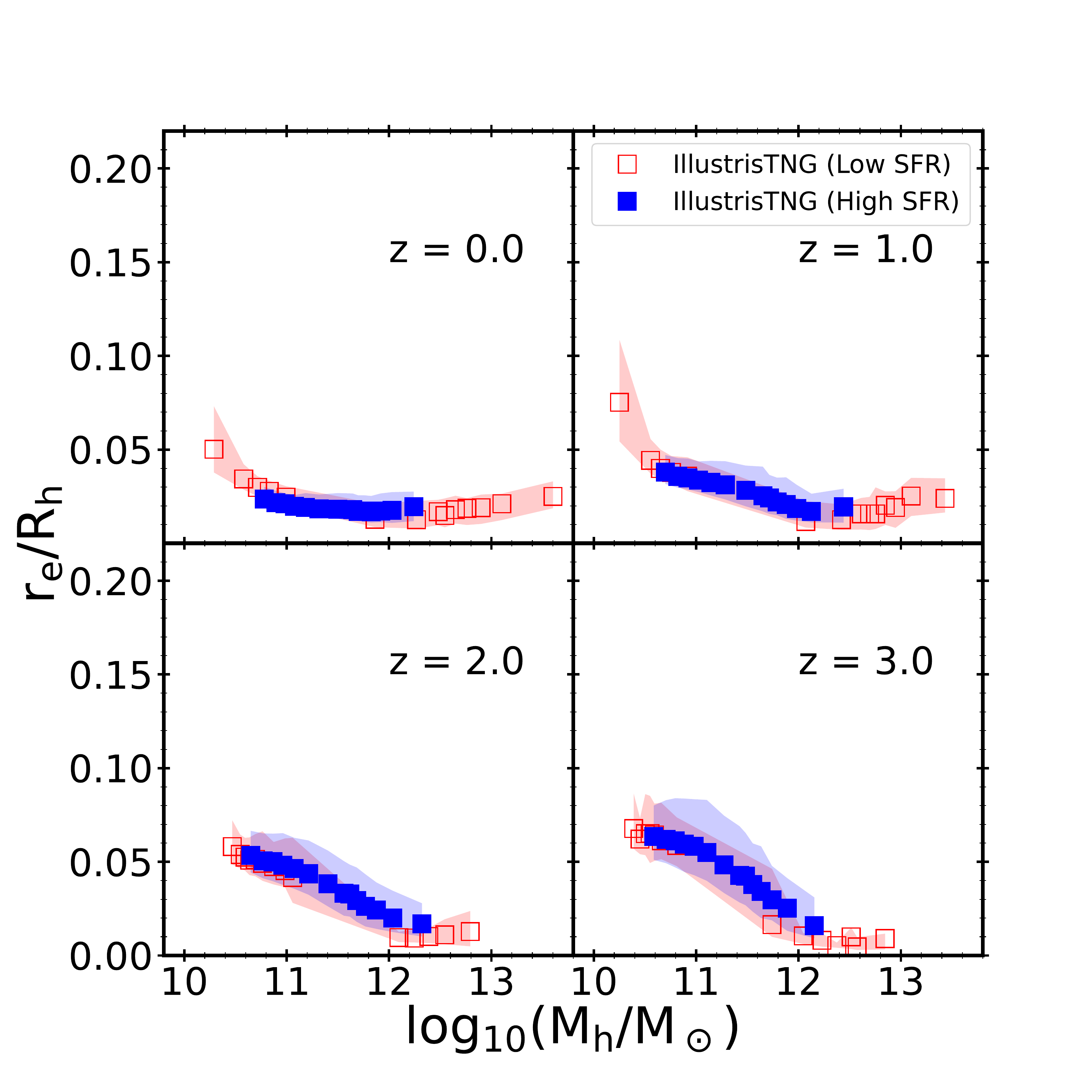}
\label{rat_s_mh_tng}
\end{subfigure}
\caption{The stellar-to-halo size ratio $r_e/R_h$ of central galaxies versus stellar mass (left) and halo mass (right). The top half presents results from the Illustris simulation, and the bottom from IllustrisTNG. The red (blue) symbols represent the medians for galaxies with low (high) SFRs, and the corresponding shaded regions represent the 16-84 percentile spread. Illustris shows a small difference between $r_e/R_h$ for low and high-SFR galaxies at the lowest masses, but there is no discernible difference between the two for IllustrisTNG.}
\label{fig:rat_s_m}
\end{figure*}

As discussed in the Introduction, it is interesting to ask whether these relations, both between stellar mass and halo mass, and that between galaxy size and halo size, differ for galaxies of different types. Here, we use star formation activity to divide galaxies into star-forming and quiescent categories, as it has been found that these two categories have distinct stellar size-mass relations \citep{vanderWelA_14a}.
To define star-forming galaxies, we follow a procedure similar to the one introduced by \citet{2015MNRAS.451.2933B}. Namely, we examine the median specific star formation rates (ssfr) of galaxies with stellar masses   $10^{9}-10^{9.5}M_\odot$ and $10^{9.5}-10^{10}M_\odot$, at a given redshift. Then, as a function of stellar mass, we define a line  by assigning these two ssfr values  to stellar masses $10^{9.25} M_\odot$ and $10^{9.75}M_\odot$ respectively. Galaxies with specific star formation rates higher than 25\% of the values denoted by the line are considered star-forming for our analysis, while the rest are considered quiescent galaxies. Such a definition results in a threshold between quenched and starforming galaxies of $10^{-10.40}/$yr at $10^{9.25}M_\odot$ and $10^{-10.45}/$yr at $10^{9.75}M_\odot$, which captures the slope of the main sequence in the simulations, and is similar to values typically used in the literature to define the quenching threshold. First, we examine the relation between stellar mass and halo mass for the two types of galaxies in Figure \ref{fig:Ms_Mh_SFR}, which shows Illustris on the left and IllustrisTNG on the right panel. We can see that the median SMHM relations are nearly indistinguishable for star-forming and quiescent galaxies in both simulations, although the dispersion is higher for quiescent galaxies. Next, in Figure \ref{fig:rat_s_m}, we show $r_e/R_h$ for star-forming and quiescent galaxies separately as a function of both stellar mass and halo mass. In the original Illustris simulation, $r_e/R_h$ is very similar for star-forming and quiescent galaxies, except that quiescent galaxies have an upturn in $r_e/R_h$ at very low halo masses (note there are no quiescent galaxies in the highest redshift bin of Illustris, $z=3$). For IllustrisTNG, $r_e/R_h$ for star-forming and quiescent galaxies are almost indistinguishable, with only a hint of a slight deviation in the lowest halo mass bin for quiescent galaxies at $z=1$ and $z=0$. 

\begin{figure*}
\hspace{0.1cm}
\begin{subfigure}{0.49\linewidth}
\includegraphics[width=\linewidth]{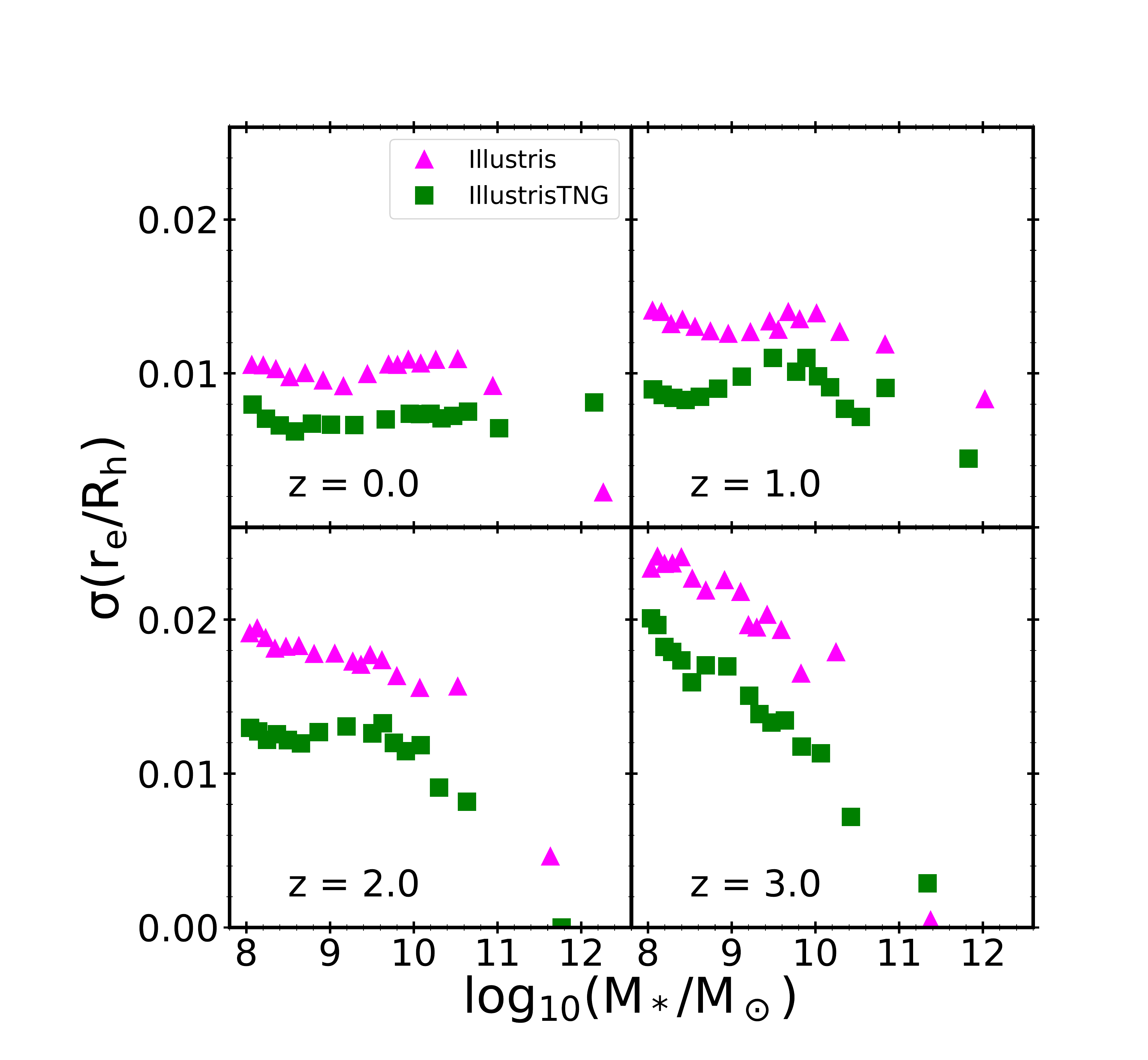}
\label{drat_ms}
\end{subfigure}
\hspace{-0.8cm}
\begin{subfigure}{0.49\linewidth}
\includegraphics[width=\linewidth]{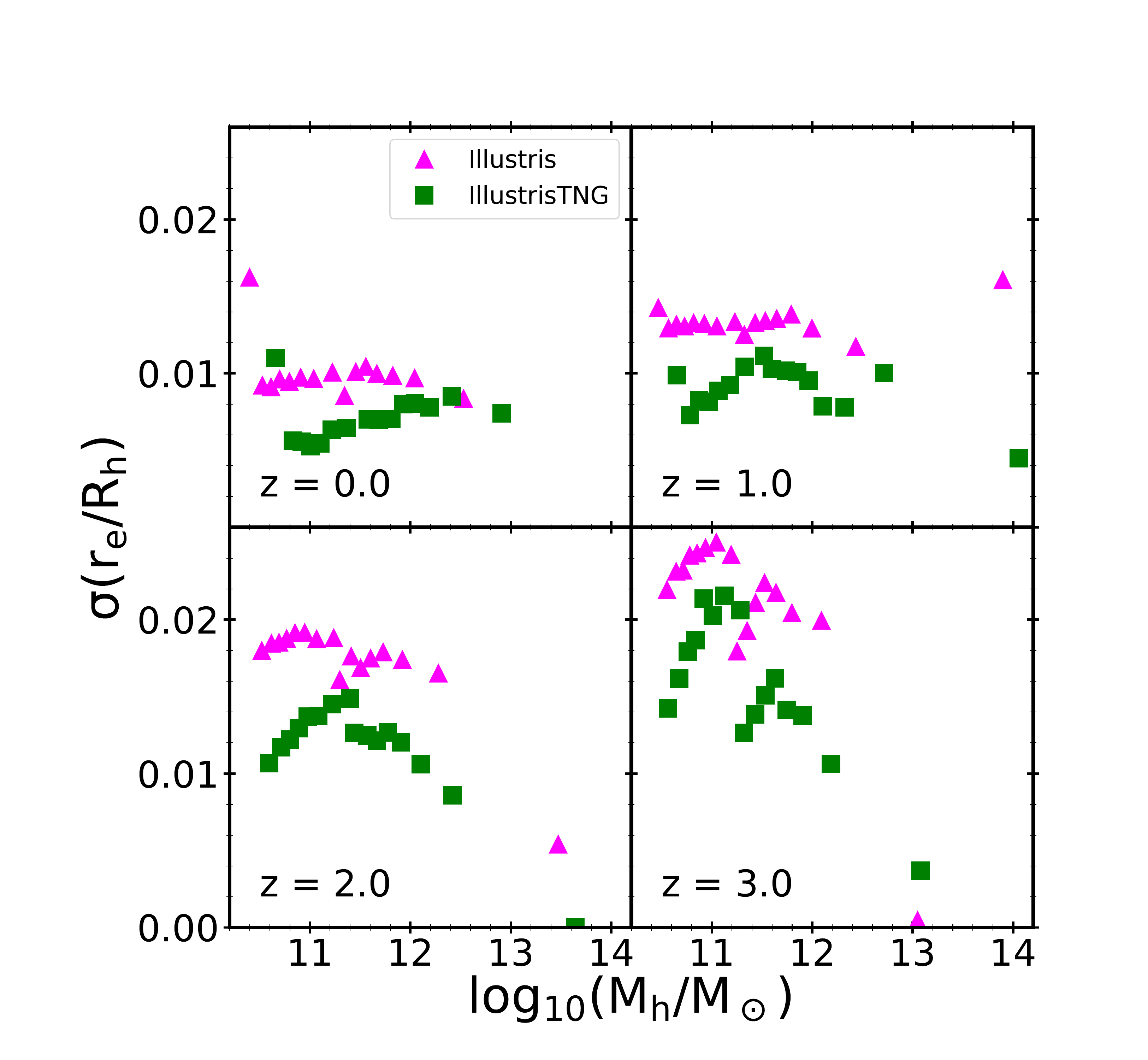}
\label{drat_mh}
\end{subfigure}

\caption{The dispersion of the stellar-to-halo size ratio as a function of stellar (halo) mass on the left (right). 
Here the dispersion is defined as half of the 84 to 16 percentile spread. Magenta (green) symbols show the results from Illustris  (IllustrisTNG). IllustrisTNG shows somewhat smaller dispersions in $r_e/R_h$ overall relative to Illustris. 
}
\label{fig:residual}
\end{figure*}

\begin{figure*}
\begin{subfigure}{0.49\linewidth}
\includegraphics[width=\columnwidth]{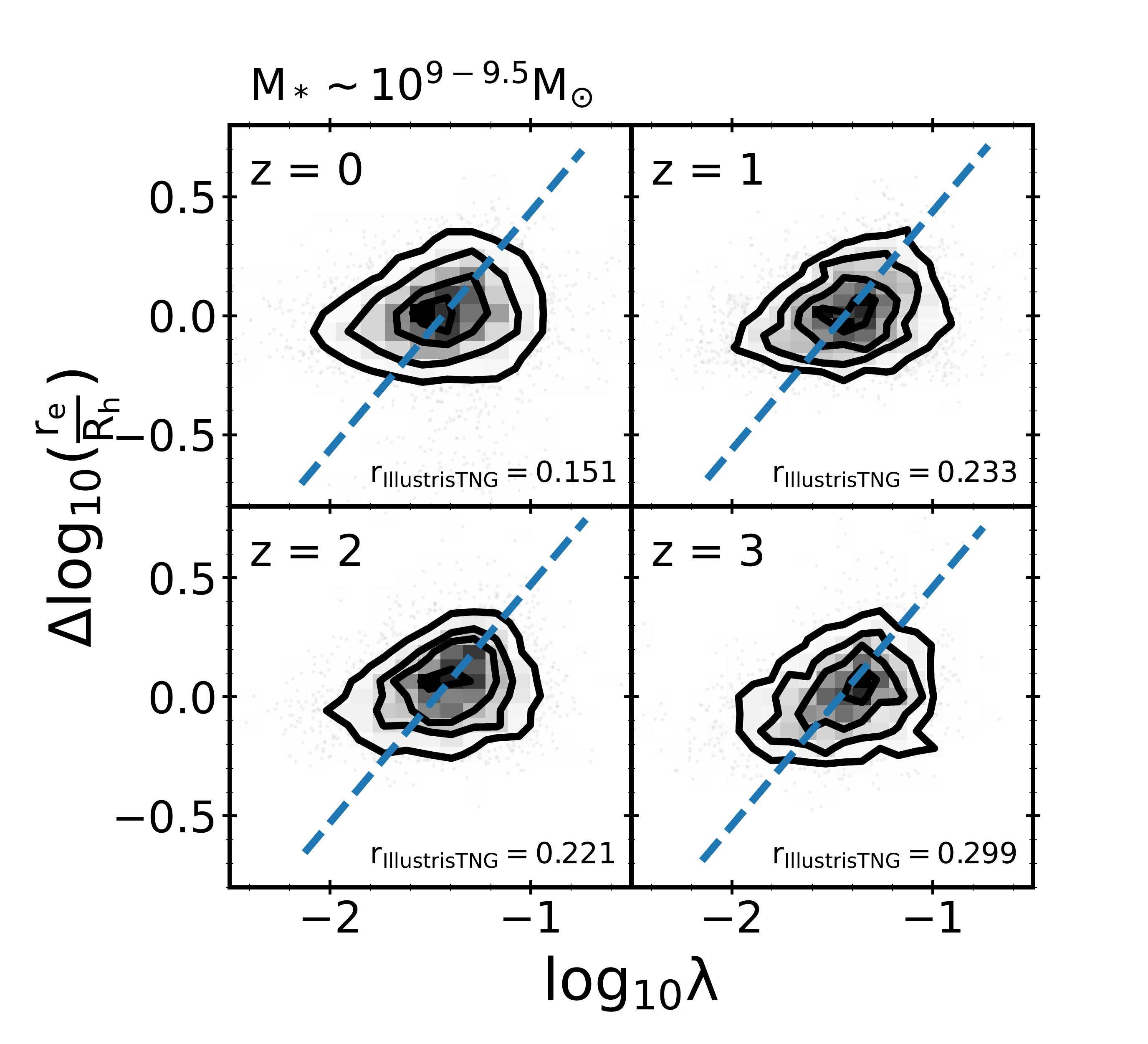}
\label{res_lamb_9_95}
\end{subfigure}
\hspace{-0.8cm}
\begin{subfigure}{0.49\linewidth}
\includegraphics[width=\linewidth]{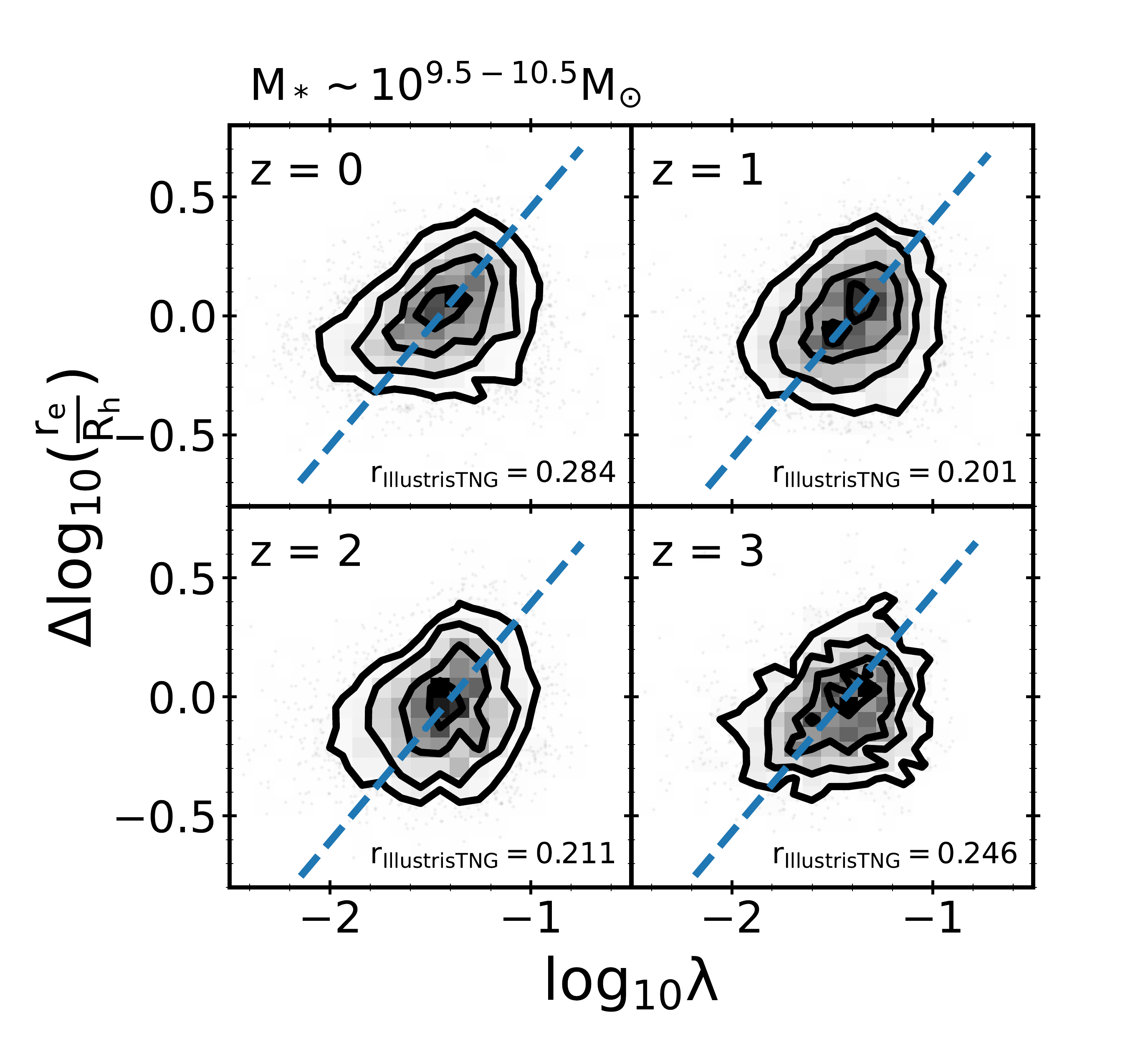}
\label{res_lamb_95_105}
\end{subfigure}
\caption{Residuals of the log stellar-to-halo size ratio $r_e/R_h$ in IllustrisTNG versus log halo spin parameter for two different stellar mass bins and four different redshifts, as indicated on the panels. The blue dashed line specifies the unit slope linear relation between these quantities.  Spearman's rank correlation coefficients are indicated on the individual panels, showing that the correlation between galaxy size and halo spin parameter in IllustrisTNG is weak. }
\label{fig:rat_lam}
\end{figure*}

\begin{figure*}
\hspace{0.1cm}

\begin{subfigure}{0.49\linewidth}
\includegraphics[width=\columnwidth]{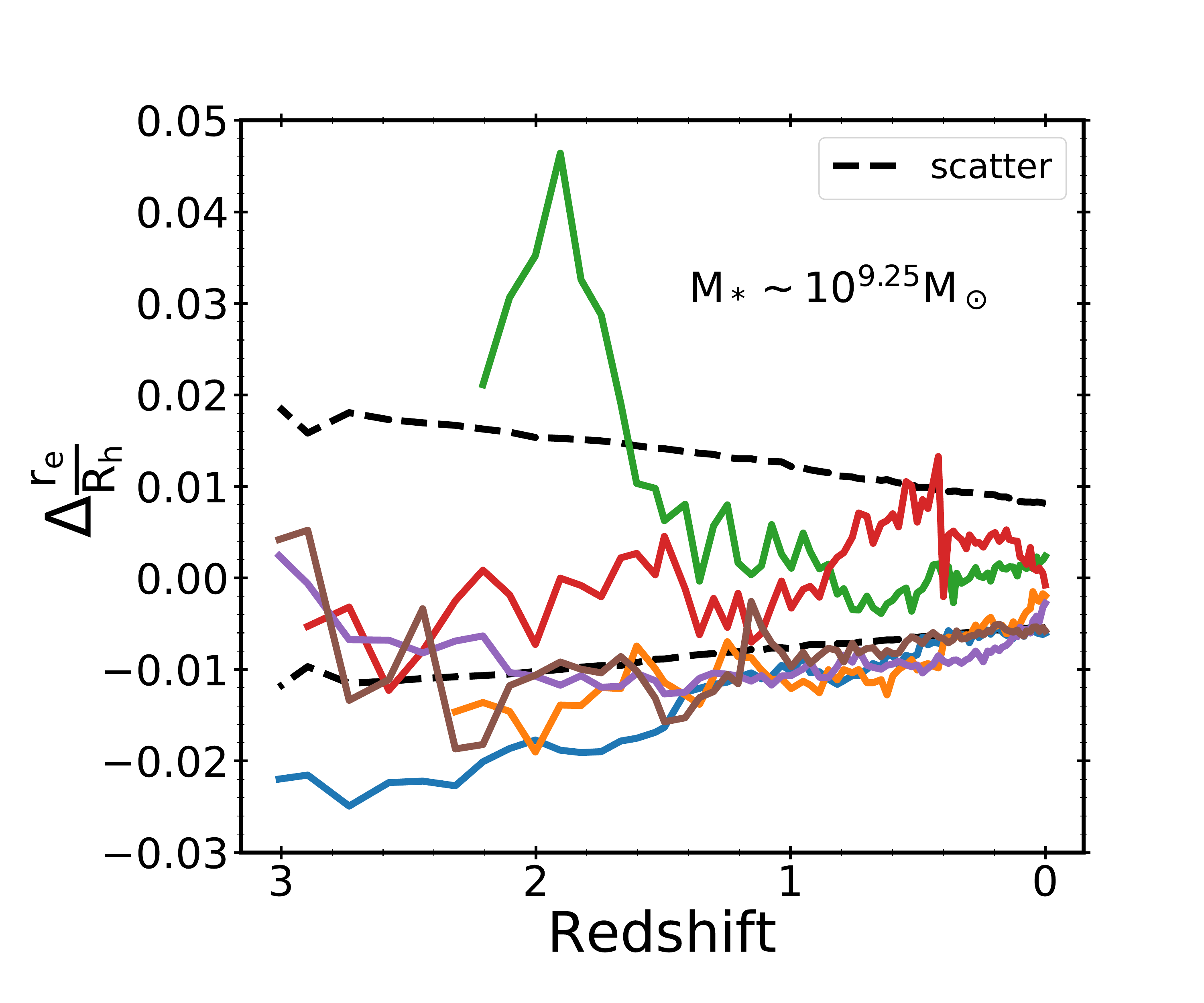}
\label{cartoon_925}
\end{subfigure}
\hspace{-0.8cm}
\begin{subfigure}{0.49\linewidth}
\includegraphics[width=\linewidth]{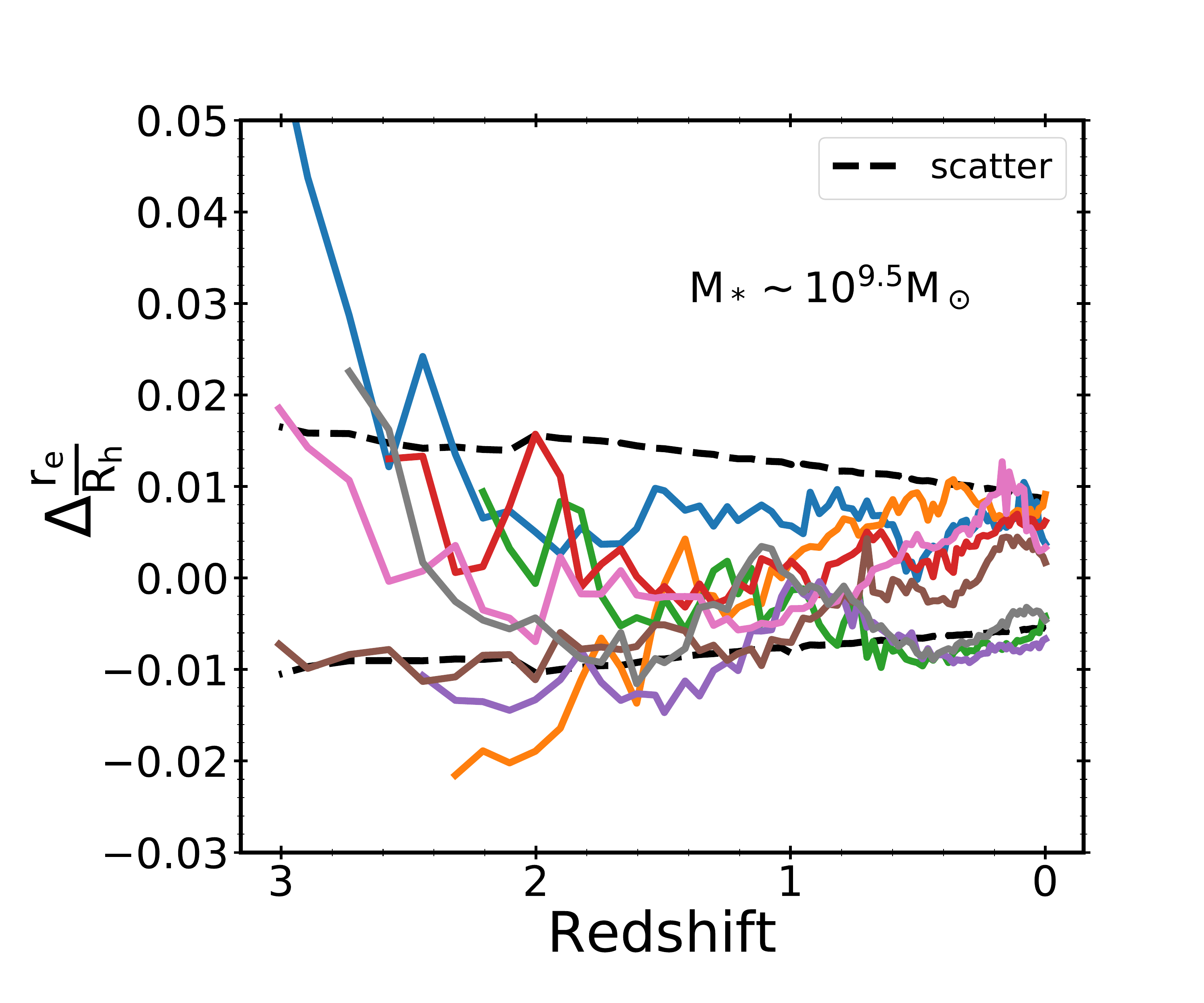}
\label{cartoon_95}
\end{subfigure}

\vspace*{-1.1cm}

\begin{subfigure}{0.49\linewidth}
\includegraphics[width=\linewidth]{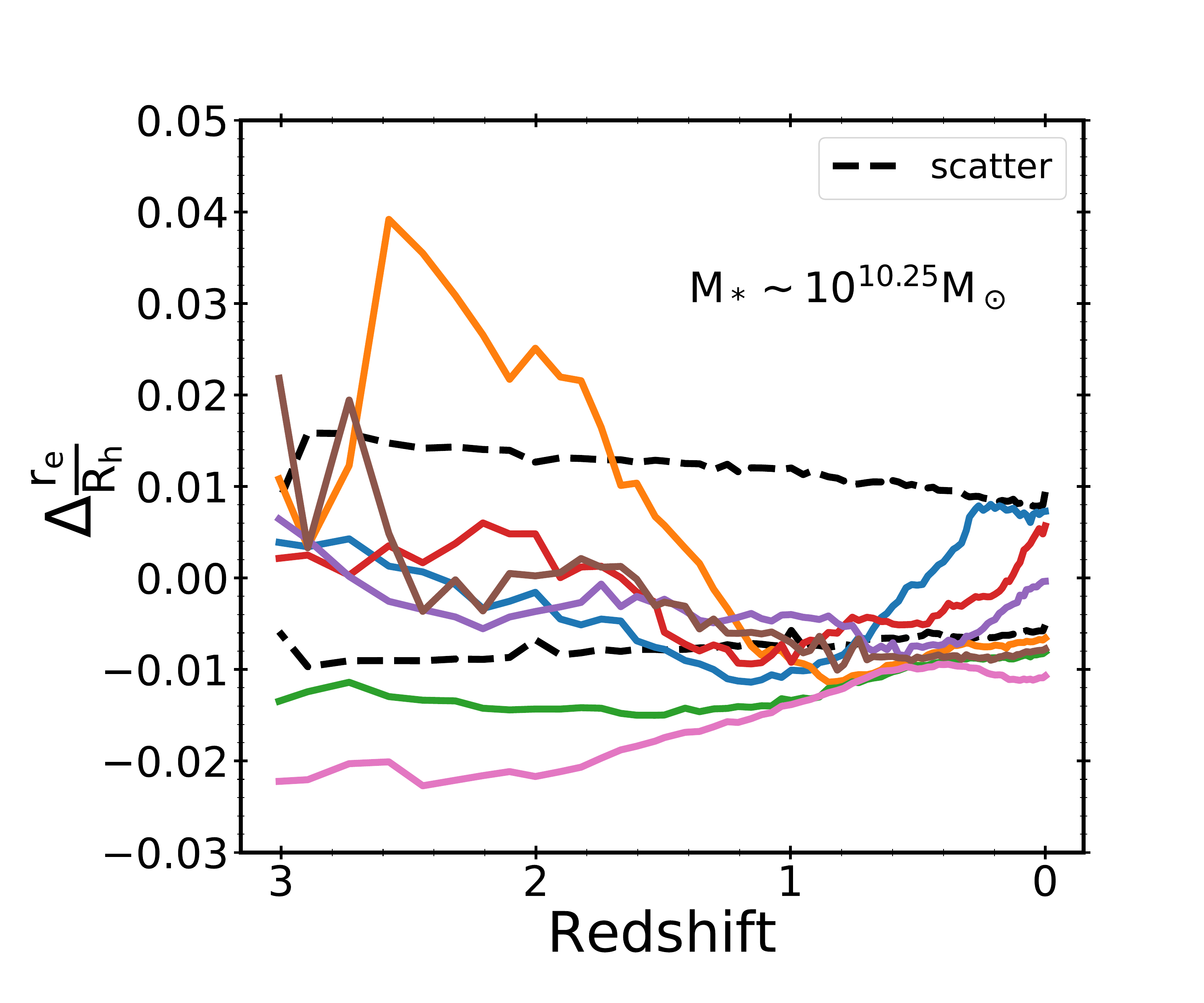}
\label{cartoon_1025}
\end{subfigure}
\hspace{-0.8cm}
\begin{subfigure}{0.49\linewidth}
\includegraphics[width=\linewidth]{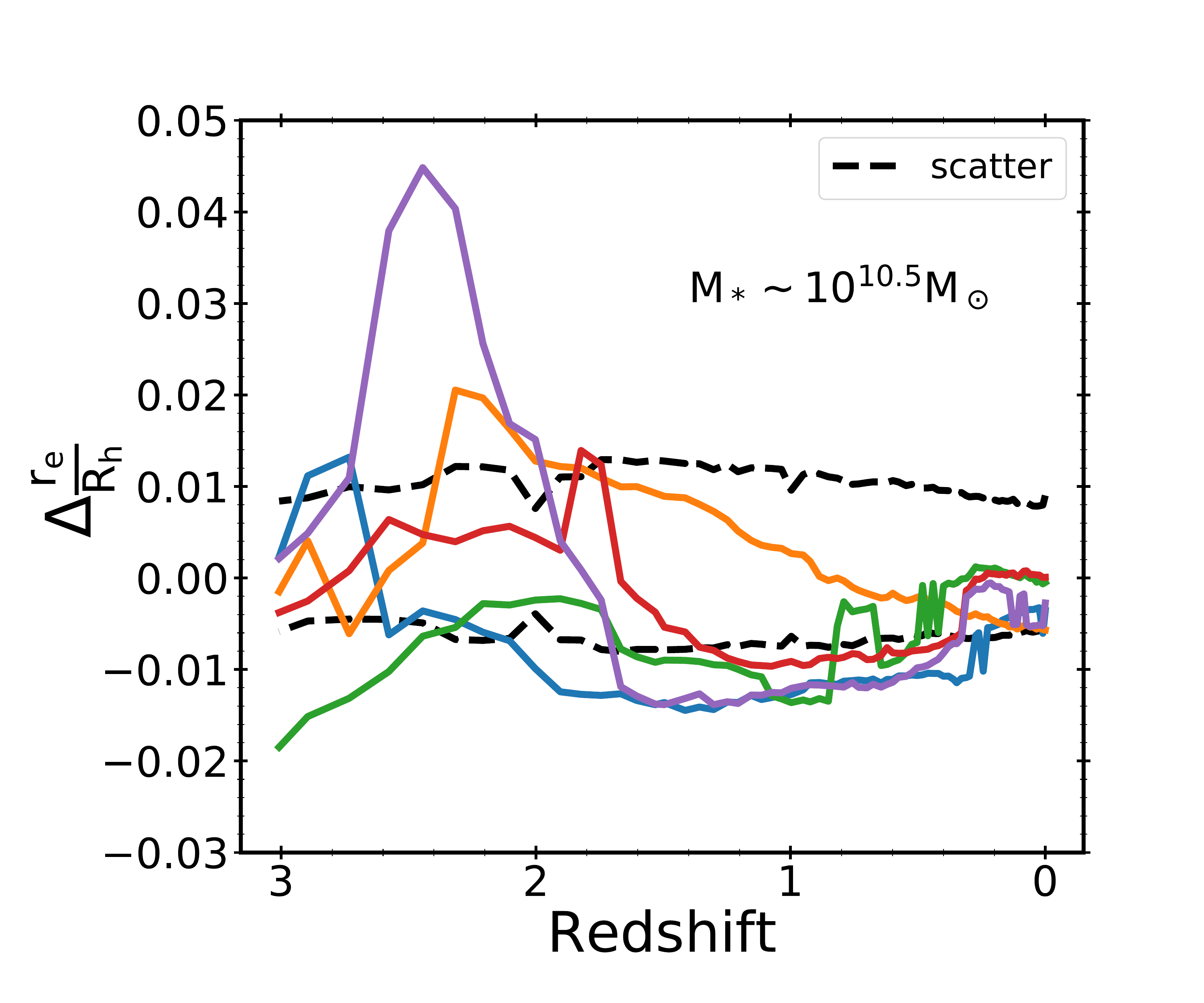}
\label{cartoon_105}
\end{subfigure}
\caption{Redshift evolution of the residual (deviation from the median) of the stellar-to-halo size ratio $r_e/R_h$ for sample galaxies with different stellar masses from IllustrisTNG. Dashed lines show the dispersion as calculated before. This illustrates that most galaxies oscillate around the median $r_e/R_h$ relation during their evolution and suggests that the dispersion in the population at a given redshift arises from these oscillations.}
\label{cartoon}
\end{figure*}

\begin{figure} 
\includegraphics[width=\columnwidth]{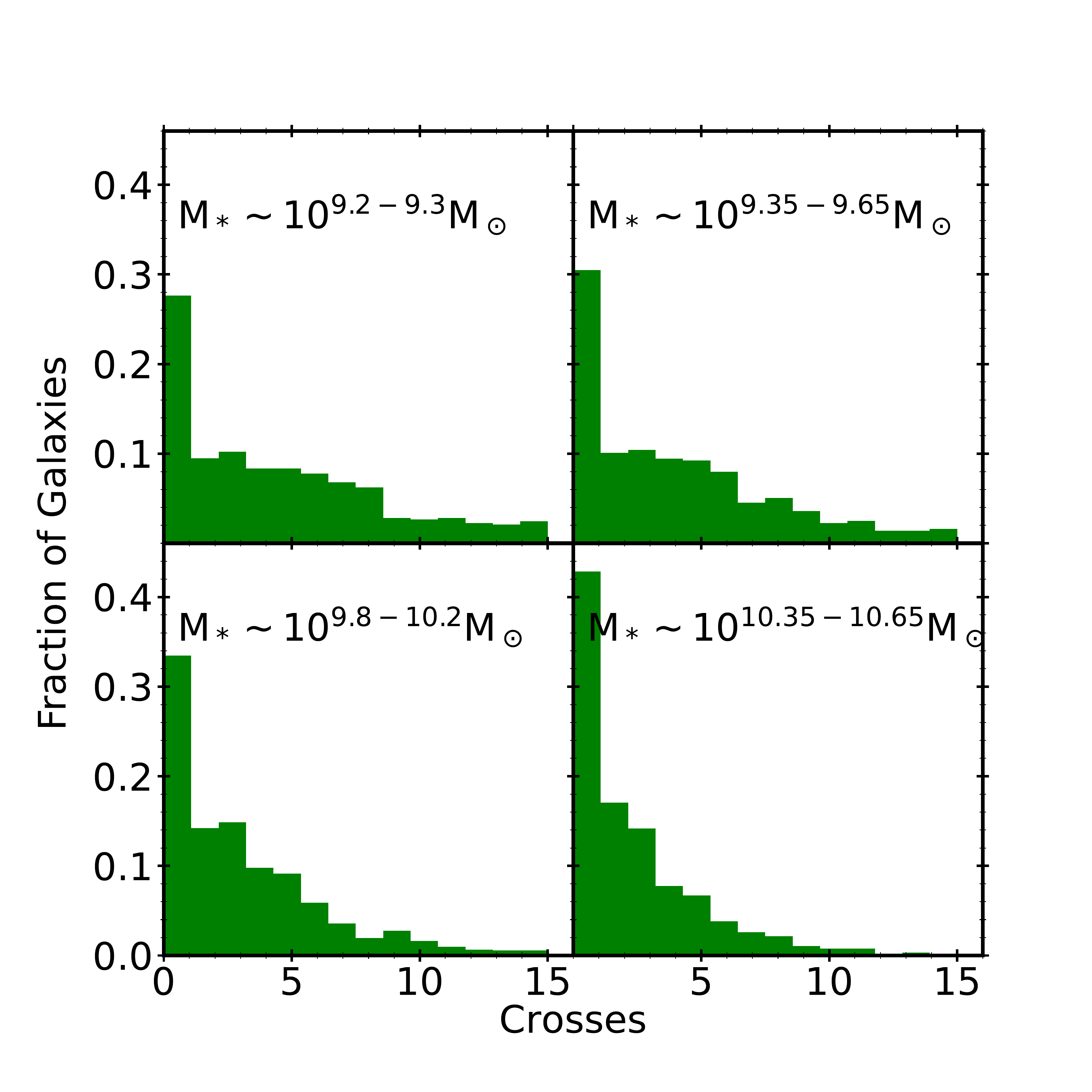}
\caption{The number of times between $z=3$ to $z=0$ that IllustrisTNG galaxies cross the median stellar-to-halo size ratio as they evolve. The indicated mass limits correspond to the galaxy masses at $z=0$. Most galaxies cross the median fewer than five times in their whole lifetime. However, between $\approx60\%-75\%$ of galaxies, depending on mass, do cross it at least once. With increasing stellar mass, the distribution becomes more skewed towards smaller values, with galaxies crossing the median less frequently.}
\label{hist_cross}
\end{figure}
\begin{figure*}
\hspace{0.1cm}
\begin{subfigure}{0.49\linewidth}
\includegraphics[width=\columnwidth,trim = {0 90 0 0}, clip]{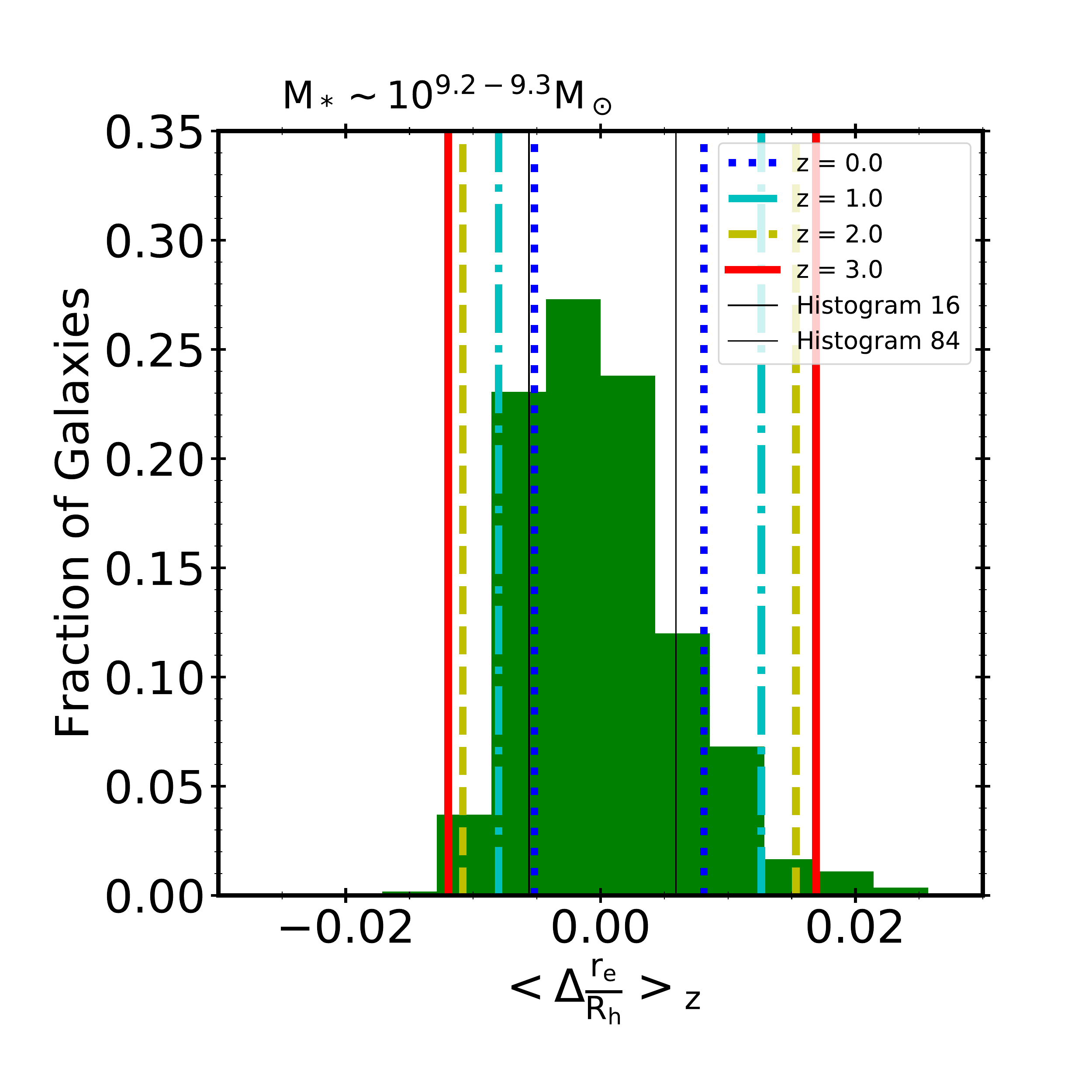}
\label{h_924}
\end{subfigure}
\hspace{-0.8cm}
\begin{subfigure}{0.49\linewidth}
\includegraphics[width=\columnwidth,trim = {0 90 0 0}, clip]{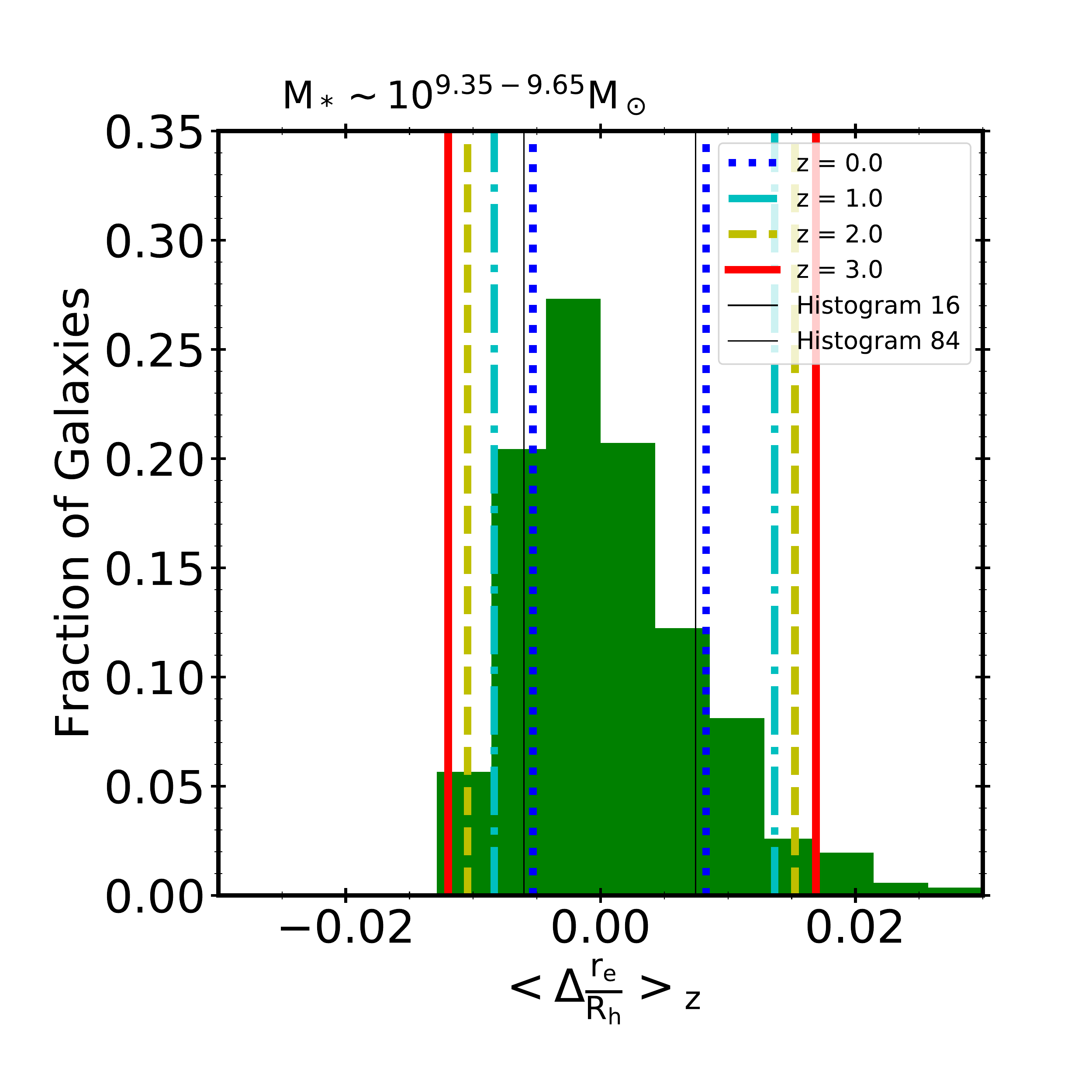}
\label{h_946}
\end{subfigure}
\vspace*{-0.6cm}

\hspace{0.1cm}
\begin{subfigure}{0.49\linewidth}
\includegraphics[width=\columnwidth]{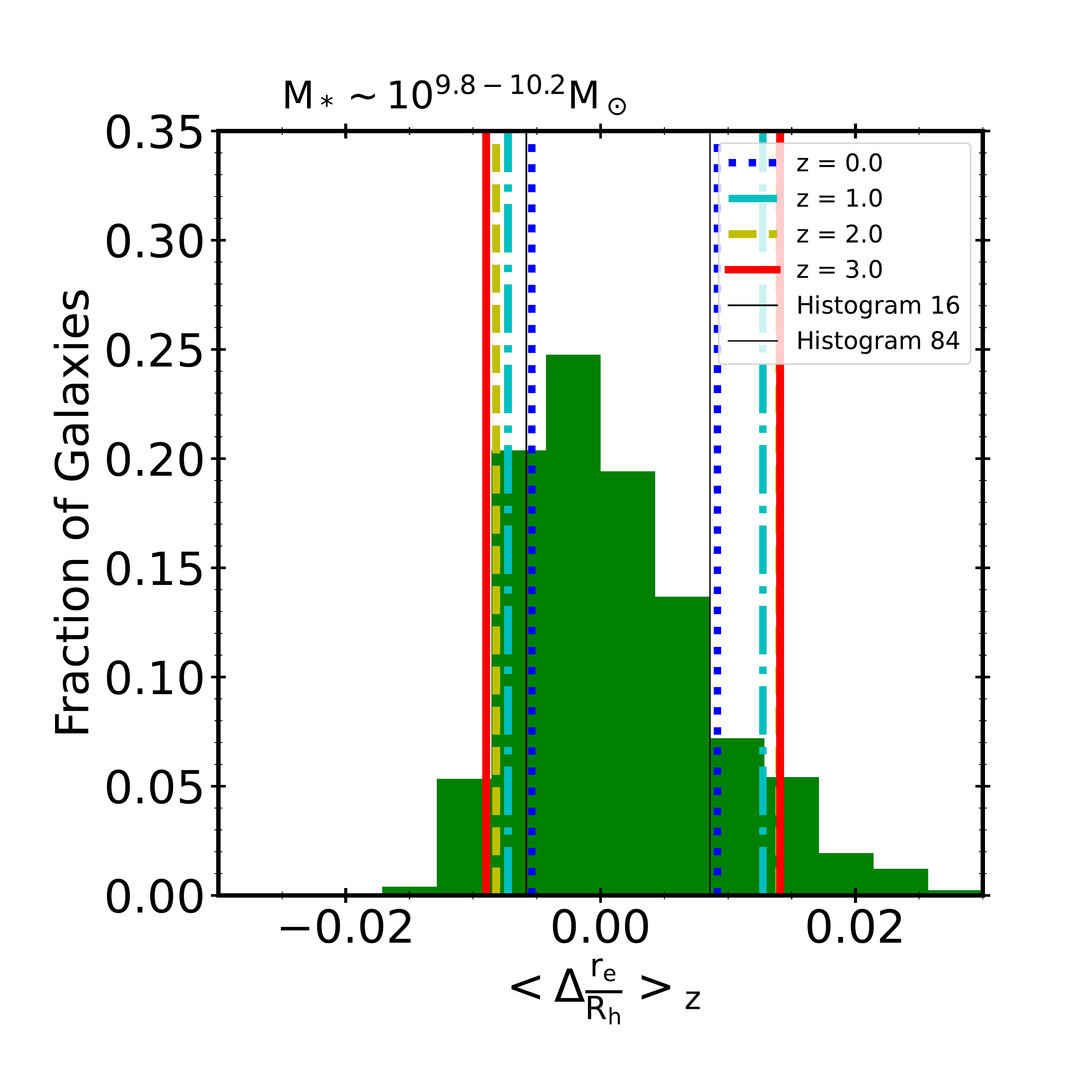}
\label{h_99}
\end{subfigure}
\hspace{-0.8cm}
\begin{subfigure}{0.49\linewidth}
\includegraphics[width=\columnwidth]{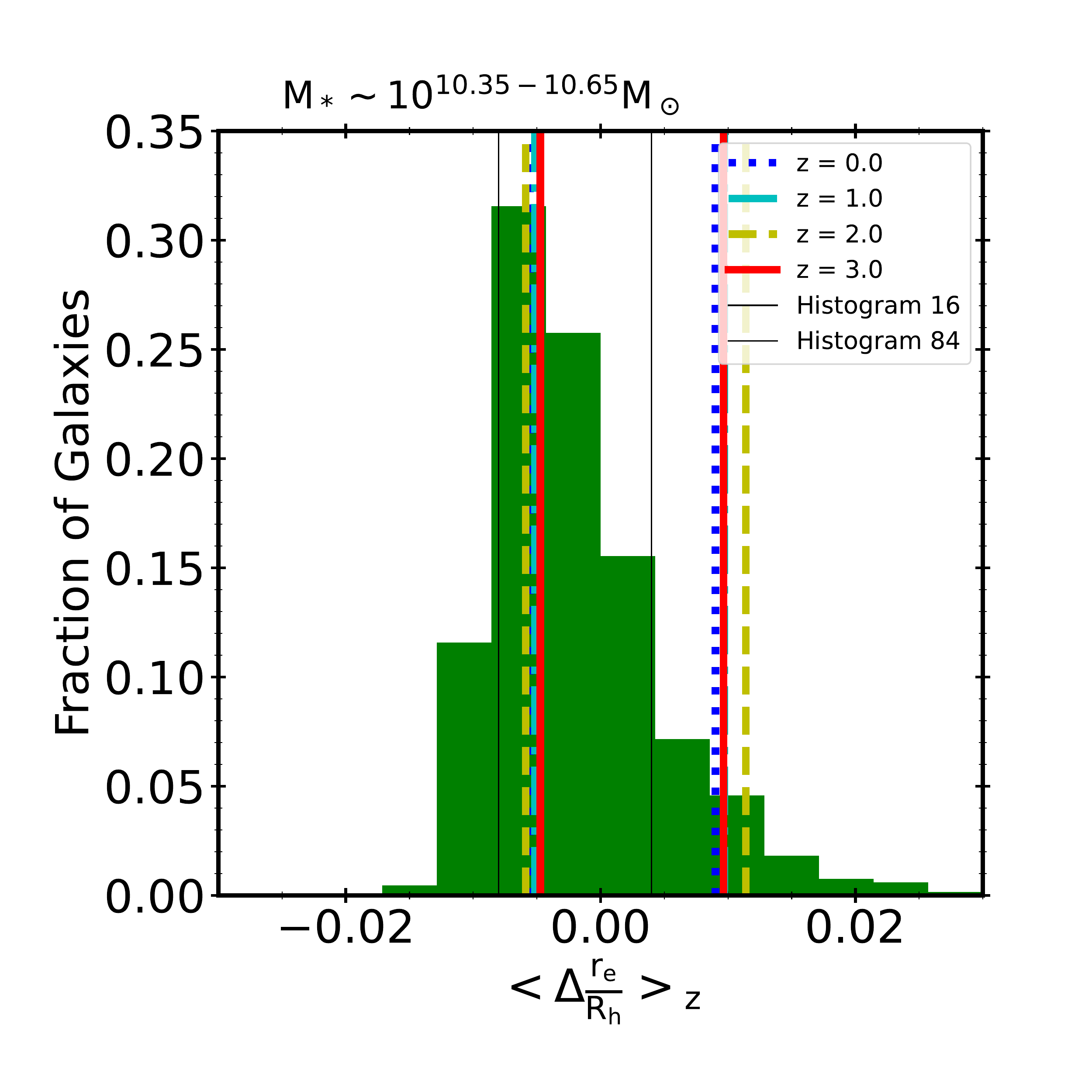}
\label{h_104}
\end{subfigure}
\caption{Histograms of the time-averaged residual of the stellar-to-halo size ratio of individual galaxies from the IllustrisTNG simulation between $z=3$ to $z=0$, with thin solid black lines representing the 16 and 84 percentiles of these histograms. The vertical lines (red, yellow, cyan, blue are for redshifts 3,2,1,0 respectively) represent the average 16 and 84 (left and right respectively) percentile spreads from Figure \ref{fig:residual}. The mass labels refer to the stellar masses at $z=0$. The peak near zero with a width less than the average scatter values indicates that the scatter in Figure \ref{fig:residual} is produced mostly due to galaxies oscillating around the median $r_e/R_h$ relation, in agreement with Figure \ref{hist_cross}.}
\label{hist_res}
\end{figure*}

\subsection{Investigation of the dispersion}

We now turn our attention to the dispersion in $r_e/R_h$ seen in both simulations in Figure~\ref{fig:rat_Mh}. The two panels in Figure \ref{fig:residual} show the scatter in $r_e/R_h$ as a function of stellar mass on the left and halo mass on the right for the Illustris and IllustrisTNG simulations. The dispersion is calculated by taking the difference between the 84 and 16 percentile values and then dividing by 2. Illustris consistently shows higher dispersions than IllustrisTNG at all redshifts. At low redshifts ($z\sim 0-1$), the dispersion is nearly constant across all stellar and halo masses.  However, the dispersion shows a decreasing trend for high mass galaxies at high redshifts. 

\par This raises the question: what is the physical origin of this dispersion in galaxy size at fixed stellar or halo mass? Is the dispersion the result of dependence on a second or higher order parameter of the halo? Or is it a result of different evolutionary paths experienced by different galaxies? As noted in the Introduction, in the classical picture of disc formation, the DM halo spin is expected to be an important second parameter in determining the galaxy size. To explore this correlation in  IllustrisTNG, we define the residual $\Delta \log_{10} (r_e/R_h)$ as the difference between the log size ratio and the median log size ratio for a given stellar mass bin and redshift. We then plot this residual against the halo spin parameter, defined as in Eqn.~\ref{eqn:spin} and calculated from the IllustrisTNG simulation itself (rather than its DM-only analog). We show this relationship in Figure \ref{fig:rat_lam}, and display the Spearman rank correlation coefficient in each panel. It is apparent that $\Delta \log_{10} (r_e/R_h)$ does not show a strong correlation with the spin parameter. For low mass galaxies, the correlation coefficient increases with redshift. For high mass galaxies, there does not seem to be a clear trend with redshift.

\par The preceding analysis shows $\Delta \log_{10} (r_e/R_h)$ at a snapshot in time for different redshifts. However, it is interesting to know whether an individual galaxy with high (or low) $\Delta \log_{10} (r_e/R_h)$ has always been high (or low) over its entire history, or do galaxies tend to frequently cross the line of zero residual, i.e., oscillate around the mean relation? In order to investigate this, we utilize merger trees that trace the main progenitor of each galaxy or halo back in time \citep{rodriguez-gomez_merger_2015}. We select galaxies from IllustrisTNG according to their $z=0$ stellar mass, and show the tracks in $\Delta \log_{10} (r_e/R_h)$ with redshift for a selection of objects in several stellar mass bins in Figure \ref{cartoon}. The residuals are calculated as the difference between the value of $r_e/R_h$ for a particular object and the median value for the same stellar mass at the same redshift. Also, for specific mass bins,  the values of the dispersion (half of the distance between the 16th and 84th percentiles) as functions of redshifts (as shown in Figure~\ref{fig:residual}) are also shown above and below the zero residual line.  As we move towards higher redshifts following the tracks, the masses of the sample galaxies tend to deviate from the specified mass bins, therefore causing a divergence in the trajectories. From this Figure, we get the impression that most galaxies tend to cross the median (i.e.~$\Delta\tfrac{r_e}{R_h}=0$) at least once over their lifetime.

To quantify this more robustly, we record the number of times that each galaxy crosses the $r_e/R_h$ median line during its lifetime between $z=3$ to $z=0$, for four different stellar mass bins in IllustrisTNG, and show the distributions of these crossing counts as histograms in Figure \ref{hist_cross}. This confirms that most galaxies ($\sim$60\%-75\%) cross the median line at least once. We also notice that the distribution becomes more skewed towards small values for higher mass bins, suggesting that massive galaxies tend not to cross the median as many times during their evolution. 

Equipped with the insights on how many times galaxies tend to oscillate around the median, we now look at how galaxies are positioned with respect to the median on average during their evolution. Figure \ref{hist_res} shows the histogram of time averaged residuals of $r_e/R_h$ for different stellar mass bins. The residuals are calculated as before, then averaged over all the snapshots. We see that in all cases, the distributions lie within the average dispersion for different redshifts. Also, the distributions are peaked around the average residual of zero. If most of the galaxies were on the same side of the median radii ratio throughout their history, we would expect to see distributions with widths comparable to the average dispersions displayed. Thus,   we infer that most of the galaxies tend to oscillate around the median several times during their evolution. Our conclusion is also supported by the findings in  Figure \ref{hist_cross}. We also see that the distributions become more and more skewed towards low values as we go to higher stellar masses, suggesting high mass galaxies with smaller sizes are more clustered near the median compared to the larger ones. It is worth noting that the vertical dispersion lines in Figure \ref{hist_res} are calculated from all the galaxies of a particular mass and redshift from the IllustrisTNG simulations. Whereas, the histograms include only galaxies that are on the main progenitor branch of a $z=0$ galaxy. However, we have verified that this subtle distinction between the two samples does not cause any significant difference in the results.

\section{Discussion}
\label{sec:discussion}
\subsection{Comparison with abundance matching results}
We have analyzed a matched set of numerical cosmological simulations, Illustris and IllustrisTNG, in which many of the physical processes are in common, but some have been altered. This provides an interesting laboratory  to study the relationship between galaxy and halo properties, and how these relationships differ in two simulations with the same underlying gravity and hydrodynamic solvers, but different implementations of physical processes. We compared the simulation predictions for the stellar to halo mass ratio $m_*/M_h$ and the galaxy size to halo size ratio $r_e/R_h$, and also compared these predictions with constraints from the structural abundance matching study of S18. We further compared both relations for galaxies divided into star-forming and quiescent types. We quantified the dispersion in $r_e/R_h$ for Illustris and IllustrisTNG, and investigated whether the residual in $r_e/R_h$ (deviation from the median) correlates with halo spin. Lastly, we explored how individual galaxies evolve in $r_e/R_h$ over their lifetimes. 

Starting with the most basic point, it has been shown before that Illustris predicts a very different relationship between galaxy mass and galaxy size than that predicted by IllustrisTNG -- the galaxy sizes are about a factor of two larger in Illustris than in IllustrisTNG \citep{Genel:2014,genel_size_2018,Pillepich:2018}. Clearly, this difference could reflect a difference either in the relationship (mean and/or dispersion) between stellar mass and halo mass, or in galaxy size and halo mass (size), or both. We have shown that the stellar mass vs. halo mass relation $m_*/M_h$ is very similar for Illustris and IllustrisTNG at all redshifts from $z\sim 0$ to 2. This means that the difference in the galaxy size-mass relation arises almost entirely from the difference in the size ratio $r_e/R_h$ --- somehow TNG haloes manage to assemble about the same mass of stars per unit total mass, but the stars are distributed very differently in radius. This is a bit surprising, as several studies have shown that varying the strength of stellar feedback often affects the slope and normalization of both $m_*/M_h$ and $r_e/R_h$. For example, \citet{Agertz:2016} showed the evolution over the lifetime of a fiducial Milky Way mass galaxy simulated with weak, fiducial, and strong feedback. They showed that weaker feedback resulted in a higher normalization and shallower slope for $m_*$-$M_h$, and a lower normalization and flatter slope for $r_e/R_h$. \citet{uebler:2014} found similar results using a different code and feedback model. 

When examining the behavior of $r_e/R_h$ as a function of halo mass for the two simulations, it is apparent that not only is the normalization of $r_e/R_h$ higher in Illustris than in IllustrisTNG at fixed $M_h$, but also that the dependence of $r_e/R_h$ on $M_h$ is much weaker in IllustrisTNG than in Illustris (Figure \ref{fig:rat_Mh}). Additionally, the difference between the normalization of $r_e/R_h$ in IllustrisTNG and Illustris at fixed mass (halo or stellar) remains approximately constant with redshift from $z\sim 3$--0. These are all important clues to the physical processes that shape these relationships. Another important clue comes from examining both $m_*/M_h$ and $r_e/R_h$ for galaxies that have been divided into star-forming and quiescent populations (which could also be viewed as a rough proxy for disc-dominated and spheroid dominated galaxies). Surprisingly, both $m_*/M_h$ and $r_e/R_h$ are nearly the same for star-forming galaxies and quiescent galaxies in \emph{both} Illustris and IllustrisTNG (with the exception of a slight upturn in $r_e/R_h$ at the lowest halo masses, $M_h \lesssim 10^{10.5} M_\odot$, where this effect is stronger in Illustris). This is surprising from two angles: first, it is well known that in observations, quiescent galaxies have different stellar mass vs. size relations than star-forming galaxies, and these relations evolve differently with redshift for the two populations. It is not well established, however, to what extent this difference seen in populations reflects different overall evolutionary tracks in individual objects vs. ``progenitor bias'' \citep[][and references therein]{vandokkum:08,carollo:2013,Keating:2015}. A related point is that using abundance matching, and assuming that all galaxies have the same SMHM relation, \citet{Huang:2017} found that $r_e/R_h$ was higher for galaxies in the highest quintile of sSFR than for those in the lowest quintile for the CANDELS sample. The second reason this result is surprising is that, from a theoretical point of view, one would expect star-forming and quiescent galaxies to evolve via different physical channels, where star-forming galaxies presumably grow mostly by accreting gas, while quiescent galaxies are thought to grow via gas poor mergers. One would expect these different channels to trace the evolution of dark matter haloes differently (we discuss this further in Section~\ref{sec:dispersion}). 

\subsection{Physical origin of differences between Illustris and IllustrisTNG}
The parameters for the IllustrisTNG simulations were tuned in light of several probes of galaxy properties, including the size-mass relation at $z=0$ \citep{Pillepich:2018}. However, the physical origin of the different predicted size-mass relations in Illustris and IllustrisTNG is not well understood, although it is thought to be primarily connected to the sub-grid treatment of stellar driven winds. In both models, these winds are implemented by randomly selecting gas particles that are converted into ``wind'' particles, which are imparted a velocity ``kick'' and temporarily decoupled from hydrodynamic forces. The primary differences between the stellar wind implementation in IllustrisTNG relative to that in Illustris are \citep{Pillepich:2018}: 1) Velocity kicks are isotropic in direction rather than parallel to the rotation axis of the galaxy 2) The wind launch velocity has an additional redshift-dependent multiplicative factor, and a floor has been imposed 3) The wind mass loading factor (which in effect determines the overall probability that a gas particle will become a wind particle) has a dependence on the gas phase metallicity, such that the mass loading is higher for lower metallicity gas. The normalization is chosen such that the mass loading for $L_*$ galaxies is similar in the two models, implying that the mass loading for lower metallicity gas is higher in IllustrisTNG than in Illustris. In addition, IllustrisTNG includes magnetic fields, which were not included in the original Illustris simulations. 

Figures 8, 10 and B1 of \citet{Pillepich:2018} show the effect of changing different pieces of the physics model in IllustrisTNG on both $m_*/M_h$ and galaxy size vs. stellar mass. Switching off magnetic fields increases $m_*/M_h$ by several tenths of dex for haloes more massive than a few times $10^{11} M_\odot$. The run with no magnetic fields leads to larger sizes at a fixed stellar mass by 20-30\% at stellar masses less than $\sim 2 \times 10^{10} M_\odot$, and up to $\sim$30 percent smaller sizes at higher stellar masses. Adopting non-isotropic winds, as in Illustris, leads to a very small change in galaxy size at fixed stellar mass relative to the TNG fiducial model. Dropping the metallicity-dependent mass loading scaling relation and removing the wind velocity floor has the largest effect on the size of any of the processes considered so far, increasing $r_e$ by a factor of about 1.7 relative to the fiducial TNG model at a stellar mass $m_* \sim 10^{10} M_\odot$. None of the changes to the IllustrisTNG model, taken individually, account for the bulk of the change to the predicted sizes relative to the original Illustris simulation. \citet{Pillepich:2018} conclude that multiple processes (at least four) interact in a non-linear way to cause the difference in galaxy size predictions.

We note that both changes 2) and 3) above will make winds stronger and more effective at removing material at high redshift, and in low mass galaxies. Na\"{\i}vely, one might have expected these changes to make galaxies \emph{even} larger (as in general, ``stronger" feedback leads to more extended discs; see above). One might also have expected a stronger redshift dependence to the emergent difference between  $r_e/R_h$ in the two models. The lesson seems to be that different implementations of stellar feedback lead to different qualitative effects on galaxy size and stellar mass, for reasons that are poorly understood. This is a topic that should be investigated systematically in the future.

\subsection{Dispersion and its origin}
\label{sec:dispersion}

Analytic and semi-analytic models of galaxy formation have commonly modeled disc sizes using the ``angular momentum partition" ansatz \citep[e.g.][]{mo_formation_1998,dutton:2007,somerville_explanation_2008,Porter:2014,henriques_galaxy_2015,lacey_unified_2016}. In this framework, it is assumed that the hot gas contained within a halo has the same spin $\lambda$ as the halo, and that the gas that forms the disc has the same specific angular momentum $j_d$ as the halo gas (or that it is related by a factor $f_j = j_d/j_h$, where $j_h$ is the specific angular momentum of the halo gas). It is then expected that the disc effective radius is given by
\begin{equation}
  r_e = \frac{1.68}{\sqrt{2}} f_j f_R(f_d, c_{\rm NFW}, \lambda)\, \lambda R_h
  \label{eqn:spin}
\end{equation}
where $f_R$ is a function with a value of order unity that accounts for the effects of an NFW \citep{navarro:1997} density profile, and for adiabatic contraction during the formation of the disc. This quantity is a weak function of the fraction of baryons in the disc ($f_d$), the NFW concentration $c_{\rm NFW}$, and the spin parameter $\lambda$ (see \citealt{somerville_explanation_2008} for details). Based on this model, we would expect the dispersion in $r_e/R_h$ to have a weak dependence on $f_d$ and $c_{\rm NFW}$, and a stronger dependence on $\lambda$. In this picture, the intrinsic dispersion in $\lambda$ for dark matter haloes, which arises from their detailed collapse/merger histories, would give rise to most of the dispersion in galaxy size at fixed halo mass. \citet{somerville_relationship_2018} show that in this picture, the dispersion in $\lambda$ found in dark matter only simulations is consistent with the observed dispersion in $m_*$ vs. $r_e$.

Several previous works have investigated the extent to which this model holds up for the galaxies produced in numerical hydrodynamic simulations. In general, these studies find no correlation or only a weak correlation between the residual in $r_e/R_h$ and the halo spin parameter measured within the virial radius \citep{Teklu:2015,zavala:2016,Zjupa:2017,desmond_galaxy-halo_2017}. \citet{zanisi_galaxy_2020} showed that the scatter in the galaxy-halo size relation for late type galaxies  could be interpreted due to the scatter in stellar angular momentum, instead of the halo spin parameter. Moreover, for late type galaxies at $z\sim 0$, \citet{zanisi_galaxy_2020} found agreement between the dispersion in galaxy sizes in IllustrisTNG and semi-empirical constraints. 
Stronger correlations have sometimes been found for sub-samples of specific types of galaxies in some simulations; for example, \citet{yang:2021} found stronger correlations between size and halo spin for Milky Way mass galaxies in the IllustrisTNG and {\sc Auriga} simulations that were selected via kinematics to be disc-dominated. However, they found weaker correlations for galaxies selected in the same way from the EAGLE and APOSTLE simulations. \citet{jiang:2019} also found a weak correlation between size and spin in the {\sc VELA} and {\sc NIHAO} zoom-in simulations, but did find a significant correlation between size and NFW halo concentration. In summary, the correlation between galaxy size and halo spin in hydrodynamic simulations seems to be dependent on the galaxy selection criteria as well as the sub-grid physics implemented in the simulations. 

For bulge-dominated galaxies, it is thought that galaxy size and mass growth are both predominantly driven by mergers. Gas-poor mergers tend to increase galaxy size, while gas-rich mergers decrease it \citep{covington:2008,naab:2009,hopkins:2009,covington:2011,oser_cosmological_2012}. Thus the dispersion in galaxy size for these objects would be expected to correlate with the number of mergers experienced by a galaxy as well as the gas content of those mergers. Covington et al. (\citeyear{covington:2008,covington:2011}) and \citet{Porter:2014} have presented results for the size evolution of bulge-dominated galaxies by implementing the size growth seen in binary merger simulations into a semi-analytic model, finding overall good agreement with observations \citep[see also][]{shankar:2013}.  To our knowledge, this has not been investigated in detail for large volume hydrodynamic simulations, but would be interesting to explore in the future.

\section{Conclusions}
\label{sec:conclusions}

We summarize our main findings and conclusions as follows: 
\begin{itemize}
    \item The relationship between stellar mass and halo mass is very similar in the Illustris and IllustrisTNG simulations from $z=3$-0. 
    \item Conversely, the relationship between galaxy radius and halo radius in Illustris and IllustrisTNG is very different. This implies that the very different galaxy stellar mass vs. size relationships in the two simulations are mainly due to the different predicted relationship between galaxy radius and halo radius. 
    \item The $r_e/R_h$ relation predicted by IllustrisTNG is in good agreement with constraints from structural abundance matching studies at low redshift. Illustris predicts a much stronger dependence of $r_e/R_h$ on halo mass, with a normalization that is higher by about a factor of two. At high redshifts, IllustrisTNG too shows a stronger $r_e/R_h$ dependence compared to the abundance matching results.
    \item Both Illustris and IllustrisTNG predict weak time evolution in $r_e/R_h$, in good agreement with the SHAM results of S18 and other similar studies.
    \item The relationships between stellar mass and halo mass, \emph{and} galaxy size and halo size, are very similar for star-forming and quiescent galaxies in both Illustris and IllustrisTNG, at all redshifts from $z=0$--3. 
    \item We quantify the dispersion in $r_e/R_h$ as a function of stellar mass, halo mass, and redshift, for both Illustris and IllustrisTNG. At $z\lesssim 1$, the dispersion is nearly constant with mass and has a value of around 0.01. At $z=2$ and above, the dispersion decreases with increasing mass, and ranges from 0.005 to 0.02. 
    \item There is only a weak correlation between the residual of $r_e/R_h$ and halo spin, at all masses and redshifts that we investigated. 
    \item When we track $r_e/R_h$ over time for the main progenitor of a galaxy, we find that most galaxies oscillate around the median value of $r_e/R_h$ several times during their lifetime, and that these oscillations are the dominant factor that gives rise to the dispersion in $r_e/R_h$.
\end{itemize}
In final conclusion, this work has continued the exploration of the link between galaxy structural properties and their haloes. We have demonstrated that quantifying the way in which galaxy properties relate to their host halo properties is a useful way to compare different simulations to one another, and also can provide useful insights into how simulations achieve or do not achieve agreement with observed scaling relations. 
	
\section*{Acknowledgements}
We would like to thank Ari Maller and the CCA galaxy formation group for valuable discussions. TK, RSS and SG acknowledge support from the Simons Foundation. TK also acknowledges support from the S.N.~Bose Scholars Program, sponsored by the Science \& Engineering Board (SERB), Department of Science and Technology (DST), Govt. of India, the Indo-U.S. Science and Technology Forum (IUSSTF) and WINStep Forward.	

\section*{Data availability}
Illustris data is available for public access at \url{https://www.illustris-project.org/data/} and IllustrisTNG data can be found at \url{https://www.tng-project.org/data/}. S18 data and data analysis scripts are available upon request to the corresponding author. 



\bibliographystyle{mnras}
\bibliography{references.bib} 



\bsp	
\label{lastpage}
\end{document}